\documentclass[11pt]{article}
\usepackage{hyperref}
\usepackage{amssymb}
\usepackage{amsmath}
\usepackage{mdframed}
\usepackage{subfig}
\usepackage{cancel}
\usepackage{enumerate}
\usepackage{color}
\usepackage{mathtools}
\usepackage{pdfpages}
\usepackage{mathrsfs}
\usepackage{enumitem}
\usepackage{mathtools}
\usepackage{graphicx}
\usepackage{mdframed}
\usepackage{amsfonts}
\usepackage{cancel}
\usepackage{placeins}

\newtheorem{definition}{Definition}[section]
\newtheorem{theorem}{Theorem}[section]
\newtheorem{assumption}{Assumption}[section]
\newtheorem{example}{Example}[section]
\newtheorem{lemma}{Lemma}[section]

\let\Item\item

\begin{document}

\centerline{{\huge Noise-induced Mixing and Multimodality}}

\medskip

\centerline{{\huge in Reaction Networks}}
 
\medskip
\bigskip

\centerline{
\renewcommand{\thefootnote}{$*$}
{\Large Tomislav Plesa\footnote{
Mathematical Institute, University of Oxford, Radcliffe Observatory 
Quarter, Woodstock Road, \\ Oxford, OX2 6GG, UK;
e-mails: plesa@maths.ox.ac.uk, erban@maths.ox.ac.uk}
\qquad 
Radek Erban$^*$
\qquad
\renewcommand{\thefootnote}{$\dagger$}
Hans G. Othmer\footnote{
School of Mathematics, University of Minnesota, 
206 Church St. SE, Minneapolis, MN 55455, USA; \\
e-mail: othmer@umn.edu}}}

\medskip
\bigskip

\noindent
{\bf Abstract}:
We analyze a class of chemical reaction networks under mass-action
kinetics and involving multiple time-scales, whose
deterministic and stochastic models display qualitative differences.
The networks are inspired by gene-regulatory networks,
and consist of a slow-subnetwork, 
describing conversions among the different gene states, 
and fast-subnetworks, 
describing biochemical interactions involving 
the gene products. 
We show that the long-term dynamics of such networks
can consist of a unique attractor at the deterministic level (unistability), 
while the long-term probability distribution at the stochastic level
may display multiple maxima (multimodality).
The dynamical differences stem from a novel phenomenon 
we call \emph{noise-induced mixing},
whereby the probability distribution of the gene products
is a linear combination of the probability
distributions of the fast-subnetworks which are 
`mixed' by the slow-subnetworks.
The results are applied in the context of systems biology,
where noise-induced mixing is shown to play
a biochemically important role, producing phenomena such
as stochastic multimodality and oscillations.

\section{Introduction} \label{sec:intro}
Biochemical processes in living systems, such as molecular 
transport, gene expression and protein synthesis,
often involve low copy-numbers of the molecular species involved.
For example, gene transcription - a process of transferring information
encoded on a DNA segment to a messenger RNA (mRNA), and
gene translation - a process by which ribosomes utilize the information
on an mRNA to produce proteins,
 typically involve interactions between 
$1$ to $3$ promoters which control transcription, on the order of ten
polymerase holoenzyme units or copies of repressor proteins, 
and on the order of a thousand RNA polymerase molecules and ribosomes~\cite{Kuthan}. 
At such low copy-numbers of some of the species, the observed dynamics of the
processes are dominated by stochastic effects, which have 
been demonstrated experimentally for single cell gene expression 
events~\cite{Spudich,Ozbudak,Levsky}. An example of this arises 
in the context of a simple pathway switch comprising two 
mutually-repressible genes, each of which produces a protein that 
inhibits expression of its antagonistic gene. Stochastic fluctuations 
present in the low copy-numbers lead to random choices of the prevailing 
pathway in a population of cells, and thus to two distinct
 phenotypes~\cite{Raj,Fraser}. Said otherwise, the probability
distribution of the phenotypes is bimodal, even in a 
genotypically-homogeneous population.
Two major sources of intrinsic noise in gene-regulatory networks are
transcriptional and translational bursting, which have been directly 
linked to DNA dynamics~\cite{Raj,Senecal,Wickramasinghe}. Transcriptional bursting 
results from slow transitions between active and inactive promoter 
states, which produces bursts of mRNA production,
while translational bursting, resulting from the random fluctuations 
in low copy-numbers of mRNA, leads to bursts in protein numbers.
 
Most signal transduction and gene-regulatory networks
 are highly interconnected, and involve numerous 
protein-protein interactions, feedback, and cross-talk 
at multiple levels. Analyzing the deterministic model of 
such complicated networks, which neglects the stochastic effects, may
be challenging on its own. Even more difficult is 
determining when there are significant differences between
the less-detailed deterministic, and the more-detailed stochastic 
models. Such differences have been called `deviant' 
in the literature~\cite{Samoilov}, and some
attempts at understanding them in terms of the underlying 
network architecture have been made~\cite{Kuwahara}, 
but there is no general understanding of when 
the deviations arise.
Of particular interest are the qualitative differences
between the long-term solutions of the deterministic 
and stochastic models~\cite{SNIC,Liao}. Central to such differences
is a relationship between multiple coexisting stable
equilibria at the deterministic level (multistability) and
coexisting maxima (modes) of the stationary probability 
distribution at the stochastic level (multimodality).
In general, mutistability and multimodality, for both transient
and long-term dynamics, do not imply each other for finite reactor 
volumes (such as in living cells)~\cite{Me2,Toth}. 
For example, even feedback-free
gene-regulatory networks, involving only first-order
reactions, which are deterministically unistable,
may be stochastically mutimodal under a suitable
time-scale separation between the gene switching 
and protein dynamics~\cite{Kepler,Andrew}. 
Long-term solutions of the deterministic model are not 
necessarily time-independent, which further complicates the analysis. 
For example, in Section~\ref{sec:examples2} we study relationships 
between a deterministic limit cycle (time-dependent long-term solution) 
and the corresponding stationary probability mass function.

The objective of this paper is to identify a 
class of chemical reaction networks
which display the `deviant' behaviours, 
and analyze the origin of such behaviours.
To this end, we consider a class of reaction networks 
with two time-scales, which consist of 
fast-subnetworks involving catalytic reactions,
and a slow-subnetwork involving conversions among the catalysts (genes). 
It is shown that a subset of such reaction networks 
are deterministically unistable, but stochastically multimodal.
We demonstrate that the cause for the observed
qualitative differences is a novel phenomenon
we call \emph{noise-induced mixing}, where
the probability distribution of
the gene products is a linear combination of 
the probability distributions of suitably
modified fast-subnetworks, which are mixed together 
by the slow-subnetworks.

The rest of the paper is organized as follows. 
In Section~\ref{sec:CRNs}, we introduce the 
mathematical background regarding chemical reaction networks.
In Section~\ref{sec:probform}, we introduce the class
of networks studied in this paper, which are 
then analyzed in Section~\ref{sec:dynamics}.
The results derived are then applied to a variety
of examples in Section~\ref{sec:applications}.
Finally, we provide summary and conclusion 
in Section~\ref{sec:conclusion}.

\section{Chemical reaction networks} \label{sec:CRNs}
In this section, chemical reaction 
networks are defined~\cite{Toth,Feinberg,Hans1,David1}, 
which are used to model the biochemical processes 
considered in this paper, together with their  
deterministic and stochastic dynamical models.
We begin with some notation.

\begin{definition} \label{definition:notation}
Set $\mathbb{R}$ is the space of real numbers,
$\mathbb{R}_{\ge}$ the space of nonnegative real numbers, and
$\mathbb{R}_{>}$ the space of positive real numbers. 
Similarly, $\mathbb{Z}$ is the space of integer numbers, 
$\mathbb{Z}_{\ge}$ the space of nonnegative integer numbers, and 
$\mathbb{Z}_{>}$ the space of positive integer numbers. 
Euclidean vectors are denoted in boldface, 
$\mathbf{x} = (x_1,x_2,\ldots, x_m) \in \mathbb{R}^m$.
The support of $\mathbf{x}$ is defined by 
$\mathrm{supp}(\mathbf{x}) = \{i \in \{1,2,\ldots, m\} | x_i \ne 0\}$.
Given a finite set $\mathcal{S}$, we denote its 
cardinality by $|\mathcal{S}|$.
\end{definition}

\begin{definition} 
A \emph{chemical reaction network} is a triple
$\{\mathcal{S}, \mathcal{C}, \mathcal{R}\}$, where
\begin{enumerate}
\item[\emph{(i)}] $\mathcal{S} = \{S_1, S_2, \ldots, S_m\}$ is the set
of species of the network.
\item[\emph{(ii)}]  $\mathcal{C}$ is the finite set of complexes of the network, 
which are nonnegative linear combinations of the species, i.e. complex 
$C \in \mathcal{C}$ reads $\sum_{i = 1}^m \nu_i S_i$, where 
$\boldsymbol{\nu} = (\nu_1, \nu_2 \ldots, \nu_m) \in \mathbb{Z}_{\ge}^m$ 
is called the stoiochiometric vector of $C$.
\item[\emph{(iii)}]  $\mathcal{R} = \{\sum_{i = 1}^m \nu_i S_i \to 
 \sum_{i = 1}^m \bar{\nu}_i S_i | \sum_{i = 1}^m \nu_i S_i, 
\sum_{i = 1}^m \bar{\nu}_i S_i \in \mathcal{C}, 
\boldsymbol{\nu} \ne \bar{\boldsymbol{\nu}}\}$ is the finite set of reactions,
with $\sum_{i = 1}^m \nu_i S_i$ and $\sum_{i = 1}^m \bar{\nu}_i S_i$
called the \emph{reactant} and \emph{product complexes}, respectively.
\end{enumerate}
\end{definition}

\noindent
For simplicity, we denote chemical reaction networks
in this paper by $\mathcal{R}$, with the species and complexes
understood in the context. Furthermore, abusing the notation 
slightly, we denote complex $\sum_{i = 1}^m \nu_i S_i$ 
by $\boldsymbol{\nu}$, when convenient.
A complex which may appear in reaction networks is 
the \emph{zero-complex}, $\boldsymbol{\nu} = \mathbf{0}$, which is
denoted by $\varnothing$ in the networks. 
Reaction $\mathbf{0} \to \bar{\boldsymbol{\nu}}$ 
then represents an inflow of the species,
while reaction $\boldsymbol{\nu} \to \mathbf{0}$ 
represents an outflow of the species~\cite{Hans1}.

The order of reaction $(\boldsymbol{\nu} \to \bar{\boldsymbol{\nu}}) \in \mathcal{R}$
is given by $\sum_{i = 1}^m \nu_i \ge 0$, while the 
\emph{order of chemical reaction network} $\mathcal{R}$ 
is then given by the order of its highest-order reaction.
We now define a 
special class of first-order networks, 
called single species complexes networks~\cite{David1} 
(also known as compartmental networks~\cite{Toth}, 
and first-order conversion networks~\cite{Hans2}), 
which play an important role in this paper.

\begin{definition} \label{definition:SSC}
First-order reaction networks $\mathcal{R}$ 
such that $(\boldsymbol{\nu} \to \bar{\boldsymbol{\nu}}) \in \mathcal{R}$
 implies $\sum_{i = 1}^m \nu_i \le 1$ and 
$\sum_{i = 1}^m \bar{\nu}_i \le 1$ are called
the \emph{single species complexes (SSC)} networks.
Such networks contain only the complexes which are either a 
single species, or the zero-complex. 
SSC networks which contain the zero-complex are said to be 
\emph{open}, otherwise they are \emph{closed}.
\end{definition}

\noindent
A reaction network $\mathcal{R}$ can be encoded as a directed graph
by identifying complexes $\mathcal{C}$ with the nodes
of the graph, and identifying each reaction
$(\boldsymbol{\nu} \to \bar{\boldsymbol{\nu}}) \in \mathcal{R}$ 
with the edge directed from the node
corresponding to $\boldsymbol{\nu}$ to the node 
corresponding to $\bar{\boldsymbol{\nu}}$. 
 A connected component of the graph is a connected subgraph which is maximal with respect to the
inclusion of edges. Each connected component is called a
\emph{linkage class}, and we denote their total number by $\ell$.

\begin{definition} \label{definition:wr}
A reaction network $\mathcal{R}$ is said to be \emph{weakly-reversible}
if the associated graph is strongly connected, i.e.
 if for any reaction 
$(\boldsymbol{\nu} \to \bar{\boldsymbol{\nu}}) \in \mathcal{R}$ 
there is a sequence of reactions, starting with a reaction 
containing $\bar{\boldsymbol{\nu}}$ as the reactant complex, and 
ending with a reaction containing $\boldsymbol{\nu}$ as the 
product complex. A reaction network is called \emph{reversible} if 
$(\boldsymbol{\nu} \to \bar{\boldsymbol{\nu}}) \in \mathcal{R}$ implies
$(\bar{\boldsymbol{\nu}} \to \boldsymbol{\nu}) \in \mathcal{R}$.
\end{definition}

\noindent
Thus, weakly-reversible networks induce a directed graph that contains
only  strongly  connected components.
When $(\boldsymbol{\nu} \to \bar{\boldsymbol{\nu}}) \in \mathcal{R}$, and 
$(\bar{\boldsymbol{\nu}} \to \boldsymbol{\nu}) \in \mathcal{R}$,
 we denote the two irreversible reactions
jointly by 
$(\boldsymbol{\nu} \xrightleftharpoons[]{} \bar{\boldsymbol{\nu}}) \in \mathcal{R}$, 
for convenience.

Before stating the last definition in this section, we define $\Delta \mathbf{x} = 
(\bar{\boldsymbol{\nu}} - \boldsymbol{\nu}) \in \mathbb{Z}_{\ge}^m$ 
to be the \emph{reaction vector} of reaction 
$(\boldsymbol{\nu} \to \bar{\boldsymbol{\nu}}) \in \mathcal{R}$. 
It quantifies the net change in
the species counts caused by a single occurrence (`firing') of the reaction. 
Set $\mathbb{S} = 
\mathrm{span}_{\{\boldsymbol{\nu} \to \bar{\boldsymbol{\nu}}
 \in \mathcal{R}\}} \{\Delta \mathbf{x} = 
(\bar{\boldsymbol{\nu}} - \boldsymbol{\nu})\}$
 is called the \emph{stoichiometric subspace}
of reaction network $\mathcal{R}$, 
 where $\mathrm{span}\{\cdot\}$ denotes 
the span of a set of vectors, 
and its dimension is denoted by $\mathrm{dim}(\mathbb{S}) = s$. 

\begin{definition} \label{definition:deficiency}
The deficiency of a reaction network $\mathcal{R}$
 is given by $\delta = |\mathcal{C}| - \ell - s$, where
$|\mathcal{C}|$ is the number of complexes, 
$\ell$ is the number of linkage classes, and 
$s$ is the dimension
of the stoichiometric subspace of network $\mathcal{R}$.
\end{definition}
Network deficiency is a nonnegative integer, $\delta \in \mathbb{Z}_{\ge}$,
which may be interpreted as the difference between the number of independent reactions
based on the reaction graph and actual number when stoichiometry is taken into
 account~\cite{Feinberg,Hans1}.
Note that SSC networks
are zero-deficient~\cite{David1}, 
which it exploited in Section~\ref{sec:stochastic}.

\subsection{The deterministic model} \label{sec:deterministic}
Let $\mathbf{x} = (x_1,x_2, \ldots, x_m) \in \mathbb{R}_{\ge}^m$
be the vector with element $x_i$ denoting the continuous 
concentration of species $S_i \in \mathcal{S}$. Furthermore, 
let us assume reactions from $\mathcal{R}$ fire according
to the \emph{deterministic mass-action kinetics}~\cite{Toth}, 
i.e. reaction $(\boldsymbol{\nu} \to 
\bar{\boldsymbol{\nu}}) \in \mathcal{R}$ fires at the rate
$k \, \mathbf{x}^{\boldsymbol{\nu}}$, where $k \in \mathbb{R}_{\ge}$
is known as the \emph{rate coefficient}, and
$\mathbf{x}^{\boldsymbol{\nu}} = \prod_{i = 1}^m x_i^{\nu_i}$,
with $0^0 = 1$.
 The deterministic model for chemical reaction network $\mathcal{R}$,
describing time-evolution of the concentration vector $\mathbf{x} = \mathbf{x}(t)$, 
where $t \in \mathbb{R}_{\ge}$ is the time-variable, 
is given by the system of autonomous first-order ordinary 
differential equations (ODEs), called the 
\emph{reaction-rate equations} (RREs)~\cite{Toth,David1}, 
which under mass-action kinetics read as
\begin{align}
\frac{\mathrm{d} \mathbf{x}}{\mathrm{d}  t} = 
\mathbf{f}(\mathbf{x}; \, \mathbf{k}) & 
=  \sum_{j = 1}^r
k_j \mathbf{x}^{\boldsymbol{\nu}_j} \Delta \mathbf{x}_j. \label{eqn:RREs}
\end{align}
Here, $|\mathcal{R}| = r$ is the total number of reactions, 
$\Delta \mathbf{x}_j = 
(\bar{\boldsymbol{\nu}}_j - \boldsymbol{\nu}_j)$ 
is the reaction vector of reaction 
$(\boldsymbol{\nu}_j \to \bar{\boldsymbol{\nu}}_j) \in \mathcal{R}$,
and $\mathbf{k} = (k_1, k_2, \ldots, k_r) \in \mathbb{R}_{\ge}^r$
is the vector of rate coefficients.
Note that, as a consequence of the mass-action kinetics, 
ODE system~(\ref{eqn:RREs}) has a polynomial
right-hand side (RHS).

A concentration vector $\mathbf{x}^* \in \mathbb{R}_{\ge}^m$,
solving~(\ref{eqn:RREs}) with the left-hand side (LHS) set to
zero, $\mathbf{f}(\mathbf{x}^*; \, \mathbf{k}) = \mathbf{0}$,
 is called an \emph{equilibrium} of the RREs. An equilibrium  $\mathbf{x}^*$
is said to be \emph{complex-balanced}~\cite{Feinberg,Hans1} if the following condition, expressing a `balancing of reactant and product complexes'
at the equilibrium, is satisfied
\begin{align}
\sum_{\{j \in \{1, 2, \ldots, r\} | \boldsymbol{\nu}_j = \mathbf{c}\}}
k_j (\mathbf{x}^*)^{\boldsymbol{\nu}_j} & = 
\sum_{\{j \in \{1, 2, \ldots, r\} | \bar{\boldsymbol{\nu}}_j = \mathbf{c}\}}
k_j (\mathbf{x}^*)^{\boldsymbol{\nu}_j}, 
\; \; \; \forall \mathbf{c} \in \mathcal{C}. \label{eq:comb}
\end{align}
For fixed rate coefficients, 
RREs which have a positive complex-balanced equilibrium are
called complex-balanced RREs. Such equations have exactly 
one positive equilibrium for each positive initial condition, 
and every such equilibrium is complex-balanced~\cite{Feinberg}. 
Furthermore, given a positive initial condition, 
the complex-balanced equilibrium is \emph{globally}
asymptotically stable, a result recently proved in~\cite{Craciun}.
 Any equilibrium on the boundary of $\mathbb{R}_{\ge}^m$
is thus unstable, so that complex-balanced RREs have a unique 
stable equilibrium for each initial condition, i.e. they are 
\emph{unistable}. We conclude this section by stating 
a theorem which relates weak-reversibility, deficiency and 
complex-balanced equilibria.

\begin{theorem} \label{theorem:comb}
{\rm (Feinberg~\cite{Feinberg})}
Let $\mathcal{R}$ be a chemical reaction network under 
mass-action kinetics. If the network is zero-deficient, 
$\delta = 0$, then the underlying {\rm RREs} have a 
positive complex-balanced equilibrium 
$\mathbf{x}^* \in \mathbb{R}_{>}^m$
if and only if network $\mathcal{R}$ is weakly-reversible.
\end{theorem}

\subsection{The stochastic model} \label{sec:stochastic}
With a slight abuse of notation, we also use
$\mathbf{x} = (x_1, x_2, \ldots, x_m) \in \mathbb{Z}_{\ge}^m$
to denote the state vector of the stochastic model, where
element $x_i$ now denotes the discrete 
copy-number of species $S_i \in \mathcal{S}$. Furthermore, 
assume reactions from $\mathcal{R}$ fire according
to the \emph{stochastic mass-action kinetics}~\cite{Hans1,David1}, 
i.e. reaction $(\boldsymbol{\nu} \to 
\bar{\boldsymbol{\nu}}) \in \mathcal{R}$ fires with propensity
(intensity) $k \, \mathbf{x}^{\underline{\boldsymbol{\nu}}}$,
 where $k \in \mathbb{R}_{\ge}$ is the rate coefficient, and
$\mathbf{x}^{\underline{\boldsymbol{\nu}}} = 
\prod_{i = 1}^m x_i^{\underline{\nu_i}}$,
where $x_i^{\underline{\nu_i}}$ is the $\nu_i$th
factorial power of $x_i$: 
$x_i^{\underline{\nu_i}} = x_i (x_i - 1) \ldots (x_i - \nu_i + 1)$
for $\nu_i > 0$, and $x_i^{\underline{0}} = 1$ for 
$x_i \in \mathbb{Z}_{\ge}$.
Let $p(\mathbf{x}, t)$ be the probability mass function (PMF), 
i.e. the probability that the copy-number vector at time 
$t \in \mathbb{R}_{\ge}$ is given by $\mathbf{x} \in \mathbb{Z}_{\ge}^m$.
The stochastic model for
chemical reaction network $\mathcal{R}$, 
describing the time-evolution of the PMF $p(\mathbf{x}, t)$, 
is given by the partial difference-differential equation, 
called the \emph{chemical master equation} 
(CME)~\cite{vanKampen,Hans1}, which under mass-action 
kinetics reads as
\begin{equation}
\frac{\partial}{\partial t} p(\mathbf{x},t)  
= 
\mathcal{L} p(\mathbf{x},t) 
= 
\sum_{j = 1}^r (E_{\mathbf{x}}^{-\Delta \mathbf{\mathbf{x}}_j} - 1) 
\big[
k \, \mathbf{x}^{\underline{\boldsymbol{\nu}}_j} p(\mathbf{x},t) 
\big], 
\label{eqn:CME}
\end{equation}
where, as in the deterministic setting, $|\mathcal{R}| = r$
 is the total number of reactions. Here, the \emph{shift-operator} 
$E_{\mathbf{x}}^{-\Delta \mathbf{\mathbf{x}}} 
=
\prod_{i = 1}^m E_{x_i}^{-\Delta x_i}$ is such 
that 
$
E_{\mathbf{x}}^{-\Delta \mathbf{\mathbf{x}}} 
[ p(\mathbf{x},t) ]
= p(\mathbf{x} - \Delta \mathbf{x},t)$, while the
linear difference operator $\mathcal{L}$ is called the \emph{forward operator}.

Function $p(\mathbf{x})$, solving~(\ref{eqn:CME}) with the LHS set to zero, 
$\mathcal{L} p(\mathbf{x}) = 0$, is called the \emph{stationary} PMF,
and it describes the stochastic behaviour of chemical reaction networks
in the long-run. It exists and is unique for the
 reaction networks considered in this paper. In general, 
the stationary PMF cannot be obtained analytically
(but computational algorithms for doing so 
are known~\cite{Liao,Hans3,Cotter,Cotter2}).
 However, in the special
case when the underlying RREs are complex-balanced, 
the stationary PMF can be obtained analytically. 
Before stating the precise result, let us note that
the stationary solution of~(\ref{eqn:CME}) may be written as~\cite{David2}
\begin{align}
p(\mathbf{x}) & = \sum_{\Gamma} a_{\Gamma} \, p_{\Gamma}(\mathbf{x}),
\label{eq:PMFg}
\end{align}
where $\{\Gamma\}$ are closed and irreducible subsets of the state-space,
$a_{\Gamma} \ge 0$, $\sum_{\{\Gamma\}} a_{\Gamma} = 1$,
and $p_{\Gamma}(\mathbf{x})$ is the unique stationary PMF
on the subset $\Gamma$, satisfying $p_{\Gamma}(\Gamma) = 1$.

\begin{theorem} \label{theorem:productform}
{\rm (Anderson, Craciun, Kurtz \cite{David2})}
Let $\mathcal{R}$ be a chemical reaction network under 
mass-action kinetics, with the rate coefficient vector 
fixed to $\mathbf{k}$ in both the {\rm RREs} and the {\rm CME}. 
Assume the underlying {\rm RREs} have a positive complex-balanced 
equilibrium $\mathbf{x}^* \in \mathbb{R}_{>}^m$. 
Then, the stationary {\rm PMF}
of the underlying {\rm CME}, given by~{\rm (\ref{eq:PMFg})},
 consists of the \emph{product-form} functions
\begin{align}
p_{\Gamma}(\mathbf{x}) & = A_{\Gamma} 
\frac{(\mathbf{x}^*)^{\mathbf{x}}}{\mathbf{x}!}, 
\; \; \; \forall \mathbf{x} \in \Gamma, \label{eq:productform}
\end{align}
and $p_{\Gamma}(\mathbf{x}) = 0$ otherwise,
where $\mathbf{x}! = x_1! \, x_2! \ldots x_m!$, 
and $A_{\Gamma} > 0$ is a normalizing constant.
\end{theorem}

\noindent
Note that Theorem~\ref{theorem:productform} is applicable for 
\emph{any} choice of rate coefficients
with $\mathrm{supp}(\mathbf{k})$ fixed, provided a reaction
network is both zero-deficient and weakly-reversible, 
by Theorem~\ref{theorem:comb}. In this paper, we
utilize two specific instances of Theorem~\ref{theorem:productform}.

\medskip

\noindent
\emph{State-space: $\Gamma = \mathbb{Z}_{\ge}^m$}.
If the state-space is given by all nonnegative integers, and it is
irreducible, then Theorem~\ref{theorem:productform} implies that 
the stationary PMF is given by the Poissonian product-form
\begin{align}
p(\mathbf{x}) & = \prod_{i = 1}^m \mathcal{P}(x_i; \, x_i^*),
\; \; \; \forall \mathbf{x} \in \mathbb{Z}_{\ge}^m, \label{eq:Poissonform}
\end{align}
where $\mathcal{P}(x_i; \, x_i^*)$ is the Poissonian with parameter $x_i^*$,
\begin{align}
\mathcal{P}(x_i; \, x_i^*) & = \exp(-x_i^*) \frac{(x_i^*)^{x_i}}{x_i!}. \nonumber
\end{align}
If an open SSC network is weakly-reversible, then the
underlying stationary PMF is of the 
form~(\ref{eq:Poissonform})~\cite{David2,Hans2}.

\medskip

\noindent
\emph{State-space: $\Gamma = \pi_m^N$}.
If the state-space is given by set
$\pi_m^N = \{\mathbf{x} = (x_1, x_2, \ldots, x_m) 
\in \mathbb{Z}_{\ge}^m | \sum_{i=1}^m x_i = N\} 
\subset \mathbb{Z}_{\ge}^m$, where $N \in 
\mathbb{Z}_{>}$, and if the set is irreducible, 
then Theorem~\ref{theorem:productform} implies that 
the stationary PMF is given by the multinomial product-form
\begin{align}
p(\mathbf{x}) & = N! \frac{(\mathbf{x}^*)^{\mathbf{x}}}{\mathbf{x}!},
\; \; \; \forall \mathbf{x} \in \pi_m^N. \label{eq:mulinomialform}
\end{align}
Here, $\mathbf{x}^*$ is the unique positive 
complex-balanced equilibrium normalized according to
$\sum_{i = 1}^m x_i^* = M = 1$, i.e. the deterministic 
conservation constant, which we denote by $M \in \mathbb{R}_{>}$,
 is set to unity.
If a closed SSC network is weakly-reversible, then the
underlying stationary PMF is of the 
form~(\ref{eq:mulinomialform})~\cite{David2,Hans2}.

\section{Fast-slow catalytic reaction networks} \label{sec:probform}
In this section, we introduce a class of chemical reaction 
networks central to this paper. Before doing so, 
let us briefly adapt the generic notation from Section~\ref{sec:CRNs}
to the specific networks studied in this section.
In what follows, the set of species is partitioned according to 
$\mathcal{S} = \mathcal{P} \cup \mathcal{G}$, 
where $\mathcal{P} = \{P_1, P_2, \ldots, P_m\}$ (`proteins')
and $\mathcal{G} = \{G_1, G_2, \ldots, G_n\}$ (`genes'). 
We also suitably partition the set of reactions, and denote 
the rate coefficients appearing in a subnetwork using the
same letter as the network subscript. For example, assuming 
network $\mathcal{R}_{\alpha}$ has $r$ reactions, 
vector $\boldsymbol{\alpha} = (\alpha_1, \alpha_2, \ldots, \alpha_r) 
\in \mathbb{R}_{\ge}^r$ contains the rate 
coefficients $\alpha_j$ appearing in the network, 
and ordered in a particular way. 
Let us stress that we allow rate coefficients, introduced 
in Sections~\ref{sec:deterministic} and~\ref{sec:stochastic}, to be 
\emph{nonnegative}, $\mathbf{k} \in \mathbb{R}_{\ge}^m$. 
In the degenerate case when a rate
coefficient is set to zero, we take the convention that the corresponding 
reaction is deleted (`switched-off') from the network, 
so that a new reaction network is obtained. For this reason,
structural properties of reaction
networks (such as those introduced in 
Definitions~\ref{definition:wr} and~\ref{definition:deficiency}),
 are stated for $\mathbf{k}$ with a fixed support.
Finally, when convenient, dependence of a reaction network on species of
interest is indicated, e.g. to emphasize that $\mathcal{R}$
involves species $\mathcal{P}$, we write 
$\mathcal{R} = \mathcal{R}(\mathcal{P})$.

\begin{definition} \label{def:intronet}
Consider mass-action reaction networks 
$\mathcal{R} = \mathcal{R}(\mathcal{P}, \mathcal{G})$,
depending on $m$ biochemical species 
$\mathcal{P} = (P_1, P_2, \ldots, P_m)$,
and 
$n$ \emph{catalytic} species 
$\mathcal{G} = (G_1, G_2, \ldots, G_n)$,
taking the following form
\begin{equation}
\mathcal{R}(\mathcal{P}, \mathcal{G}) = 
\mathcal{R}_{\alpha,\beta}(\mathcal{P}, \mathcal{G}) 
\cup 
\mathcal{R}_{\gamma}^{\varepsilon}(\mathcal{P},\mathcal{G}), \label{eq:R}
\end{equation}
with
\begin{equation}
\mathcal{R}_{\alpha,\beta}(\mathcal{P}, \mathcal{G}) = 
\mathcal{R}_{\alpha}(\mathcal{P}; \,\mathcal{G})
\cup \mathcal{R}_{\beta}(\mathcal{P}). \label{eq:Rab}
\end{equation}
All the reactions in 
$\mathcal{R}_{\alpha} = \mathcal{R}_{\alpha}(\mathcal{P}; \,\mathcal{G})$ 
are catalysed by (a subset of) catalysts $\mathcal{G}$,
and the network is called the \emph{catalysed network}.
On the other hand, all the reactions in 
$\mathcal{R}_{\beta} = \mathcal{R}_{\beta}(\mathcal{P})$ are independent of the
catalysts $\mathcal{G}$, and the network is called the
\emph{uncatalysed network}.
Network $\mathcal{R}_{\gamma}^{\varepsilon} = 
\mathcal{R}_{\gamma}^{\varepsilon}(\mathcal{P},\mathcal{G})$ 
is called the \emph{catalysing network}.
If all the reactions in the catalysing network
depend only on the catalysts $\mathcal{G}$,
then the network is said to be \emph{unregulated}, otherwise,
if at least one reaction depends on some of the species $\mathcal{P}$,
the network is said to be \emph{regulated}.

Let network $\mathcal{R}_{\delta} = \mathcal{R}_{\delta}(\mathcal{P})$,
obtained by removing the catalysts $\mathcal{G}$
from the reactions underlying $\mathcal{R}_{\alpha}$,
be called the \emph{decatalysed network}.
We call catalyst-independent network
\begin{equation}
\mathcal{R}_{\delta,\beta}(\mathcal{P}) = 
\mathcal{R}_{\delta}(\mathcal{P})
\cup \mathcal{R}_{\beta}(\mathcal{P}), \label{eq:Rdb}
\end{equation}
the \emph{auxiliary network} corresponding to~{\rm (\ref{eq:R})}.
\end{definition}

\noindent
To facilitate the analysis of network~(\ref{eq:R}), we introduce 
several assumptions concerning its structure and dynamics, starting 
with assumptions about catalysed and auxiliary networks.

\begin{assumption}[\textbf{Catalysed network}]  
\label{assumption:catalysed}
Structurally, the \emph{catalysed network} 
$\mathcal{R}_{\alpha}$, given in~{\rm (\ref{eq:Rab})}
is assumed to take the following separable form
\begin{equation}
\mathcal{R}_{\alpha}(\mathcal{P}; \,\mathcal{G}) 
= 
\mathop{\bigcup}_{i = 1}^n 
\mathcal{R}_{\alpha_i}(\mathcal{P}; \, G_i), \label{eq:R0a}
\end{equation}
where 
$
\boldsymbol{\alpha} 
= 
(\boldsymbol{\alpha}_1, 
\boldsymbol{\alpha}_2 \ldots, \boldsymbol{\alpha}_n)
$,
with vector $\boldsymbol{\alpha}_i$ containing the rate coefficients 
appearing in the $i$-th catalysed network $\mathcal{R}_{\alpha_i}$.
Each subnetwork $\mathcal{R}_{\alpha_i}$
is \emph{first-order catalytic} in exactly one species $G_i$,
with the $j$-th reaction given by 
\begin{equation}
r_{i j}: 
G_i 
+ \left(\sum_{k = 1}^m \nu_{i j}^k P_k \right) 
\xrightarrow[]{\alpha_{i j}} G_i 
+ \left( \sum_{k = 1}^m \bar{\nu}_{i j}^{k} P_k 
\right),
\qquad \mbox{for} \quad i \in \{1, 2, \ldots, n\}.
\label{eq:R0ai}  
\end{equation}
Dynamically, the {\rm CME} underlying 
\emph{auxiliary network}~{\rm (\ref{eq:Rdb})},
which, considering~{\rm (\ref{eq:R0a})}, reads as
\begin{equation}
\mathcal{R}_{\delta,\beta}(\mathcal{P})
= 
\left( \mathop{\bigcup}_{i = 1}^n 
\mathcal{R}_{\delta_i}(\mathcal{P}) \right)
\cup \mathcal{R}_{\beta}(\mathcal{P}), \label{eq:R0db}
\end{equation}
is assumed to have a unique stationary {\rm PMF}
for any choice of the 
underlying rate coefficients 
$(\boldsymbol{\delta}, \boldsymbol{\beta}) = 
(\boldsymbol{\delta}_1, \boldsymbol{\delta}_2,
\ldots,\boldsymbol{\delta}_n,\boldsymbol{\beta})$,
and we call it the \emph{auxiliary PMF}.
In other words, the stochastic process induced by the network is
unconditionally ergodic.
Here, the $j$-th reaction in the
$i$-th decatalysed network $\mathcal{R}_{\delta_i}$ reads
\begin{equation}
r_{i j}: 
\sum_{k = 1}^m \nu_{i j}^k P_k 
\xrightarrow[]{\delta_{i j}} \sum_{k = 1}^m \bar{\nu}_{i j}^{k} P_k,
\qquad \mbox{for} \quad i \in \{1, 2, \ldots, n\}.
\label{eq:R0di}  
\end{equation}
\end{assumption}

\noindent
The following assumptions are made on the 
structural properties of the catalysing
network underlying~(\ref{eq:R}), 
where Definitions~\ref{definition:SSC} 
and~\ref{definition:wr} are used.

\begin{assumption}[\textbf{Catalysing network}] 
 \label{assumption:catalysing}
The catalysing network 
$\mathcal{R}_{\gamma}^{\varepsilon} = 
\mathcal{R}_{\gamma}^{\varepsilon}(\mathcal{G})$,
given in~{\rm (\ref{eq:R})}, is assumed to 
be \emph{unregulated}. Furthermore, it is assumed to be a closed
{\rm SSC} network which is weakly-reversible,
with the reactions taking the following form
\begin{equation}
r_{i j}:  
G_i 
\xrightarrow[]{\varepsilon \gamma_{i j}} G_j, 
\; \; \; \; i,j \in \{1, 2, \ldots, n\}, 
\; \; \; i \ne j. \label{eq:R1unreg}  
\end{equation}
\end{assumption}

\noindent
The final assumption involves the rate coefficients 
appearing in~(\ref{eq:R}).

\begin{assumption}[\textbf{Time-scale separation}] 
\label{assumption:fastslow}
Consider the nonnegative rate coefficient vectors 
 $\boldsymbol{\alpha} = 
(\boldsymbol{\alpha}_1, \boldsymbol{\alpha}_2
\ldots, \boldsymbol{\alpha}_n)$, 
$\boldsymbol{\beta}$ and $\varepsilon \boldsymbol{\gamma}$,
appearing in the catalysed network 
$\mathcal{R}_{\alpha} = \cup_{i=1}^n \mathcal{R}_{\alpha_i}$,
uncatalysed network $\mathcal{R}_{\beta}$
and catalysing network 
$\mathcal{R}_{\gamma}^{\varepsilon}$, respectively.
It is assumed that  
 $0 < \varepsilon \ll 1$,
while the positive elements in 
$\boldsymbol{\alpha}$, 
$\boldsymbol{\beta}$ and $\boldsymbol{\gamma}$
are of order one, $\mathcal{O}(1)$, with respect 
to $\varepsilon$.
In other words, the catalysed and uncatalysed networks,
jointly denoted $\mathcal{R}_{\alpha,\beta}$,
are \emph{fast}, while the catalysing network 
$\mathcal{R}_{\gamma}^{\varepsilon}$ is \emph{slow}.
\end{assumption}

\noindent
Network~(\ref{eq:R}), under three
Assumptions~\ref{assumption:catalysed}--\ref{assumption:fastslow}, 
describes feedback-free gene-regulatory 
networks~\cite{Kepler}. In particular, species $\mathcal{G}$ 
may be seen as different gene expressions (gene with different 
operator occupancy), while 
$\mathcal{P}$ can represent suitable gene products 
(such as mRNAs and proteins) and species which can interact 
with the products. Under this interpretation,
the unregulated catalysing network 
$\mathcal{R}_{\gamma}^{\varepsilon}(\mathcal{G})$, with 
reactions~(\ref{eq:R1unreg}), describes the gene which 
slowly switches between $n$ different states, independently 
of the gene products. Catalysed network 
$\mathcal{R}_{\alpha_i}(\mathcal{P}; \, G_i)$,
 with reactions~(\ref{eq:R0ai}),
describes the action of the gene in state $G_i$ 
on the products $\mathcal{P}$. Finally, the uncatalysed network 
$\mathcal{R}_{\beta}(\mathcal{P})$ describes interactions 
between gene products (and possibly other
molecules), such as formations of dimers and higher-order 
oligomers, which take place independently of the gene state.

\begin{example} \label{ex:fastslownet}
Consider the following fast-slow network
\begin{align}
\mathcal{R}_{\alpha_1}: \; & & G_1 & 
\xrightleftharpoons[\alpha_{1 2}]
{\alpha_{1 1}} G_1 + P_1, \nonumber \\
\mathcal{R}_{\alpha_2}: \; & & G_2 & 
\xrightarrow[]{\alpha_{2 1}} G_2 + P_1, \label{eq:fastslow1} \\
\mathcal{R}_{\beta}: \; & & P_1 & 
\xrightarrow[]{\beta_1} \varnothing, \nonumber \\
\mathcal{R}_{\gamma}^{\varepsilon}: \;
 & & G_1 & \xrightleftharpoons[\varepsilon \gamma_{2 1}]
{\varepsilon \gamma_{1 2}} G_2, 
\; \; \; \; \; 0 < \varepsilon \ll 1, \nonumber 
\end{align}
involving species $\mathcal{P} = (P_1)$ and 
catalysts $\mathcal{G} = (G_1,G_2)$,
 with rate coefficients (all assumed to be positive)
$\boldsymbol{\alpha}_1 = (\alpha_{1 1}, \alpha_{1 2})$,
$\boldsymbol{\alpha}_{2} = (\alpha_{2 1})$, 
$\boldsymbol{\beta} = (\beta_1)$, 
$\boldsymbol{\gamma} = (\gamma_{1 2}, \gamma_{2 1})$.
Here, $\xrightleftharpoons[]{}$ denotes a reversible
reaction (see also Section~{\rm \ref{sec:CRNs})}.

There are two catalysed networks of the form~{\rm (\ref{eq:R0ai})}
embedded in~{\rm (\ref{eq:fastslow1})}: network
$\mathcal{R}_{\alpha_1} = \mathcal{R}_{\alpha_1}(P_1; \, G_1)$, 
describing a production and degradation of $P_1$ catalysed by $G_1$, and 
$\mathcal{R}_{\alpha_2} = \mathcal{R}_{\alpha_2}(P_1; \, G_2)$, 
describing a production of $P_1$ catalysed by $G_2$.
The uncatalysed network, 
$\mathcal{R}_{\beta} = \mathcal{R}_{\beta}(P_1)$,
describes a degradation of $P_1$,
occurring independently of $G_1$ and $G_2$.
Finally, the unregulated catalysing network 
$\mathcal{R}_{\gamma}^{\varepsilon} = 
\mathcal{R}_{\gamma}^{\varepsilon}(\mathcal{G})$
is a closed and reversible {\rm SSC} network
of the form~{\rm (\ref{eq:R1unreg})}, with $n = 2$
(where we implicitly assume $\gamma_{1 2}, \gamma_{2 1} > 0$).
Network~{\rm (\ref{eq:fastslow1})} may be interpreted
as describing a gene slowly switching between two 
expressions $G_1$ and $G_2$. When in state $G_1$, 
the gene produces and degrades protein $P_1$,
while when in state $G_2$, it only produces $P_1$, 
but generally at a different rate than when it is in state $G_1$. 
Furthermore, $P_1$ may also spontaneously degrade.
Networks similar to~{\rm (\ref{eq:fastslow1})}
have been analysed in the literature~\cite{Kepler,Andrew}.

The auxiliary network $\mathcal{R}_{\delta,\beta}$,  
given generally by~{\rm (\ref{eq:R0db})}--{\rm (\ref{eq:R0di})},
in the specific case of network~{\rm (\ref{eq:fastslow1})} 
reads as
\begin{align}
\mathcal{R}_{\delta_1}: \; & & \varnothing & 
\xrightleftharpoons[\delta_{1 2}]
{\delta_{1 1}} P_1, \nonumber \\
\mathcal{R}_{\delta_2}: \; & & \varnothing & 
\xrightarrow[]{\delta_{2 1}} P_1, \label{eq:decat1} \\
\mathcal{R}_{\beta}: \; & & P_1 & 
\xrightarrow[]{\beta_1} \varnothing. \nonumber  
\end{align}
The auxiliary network~{\rm (\ref{eq:decat1})} is equivalent 
to 
$$
\varnothing 
\xrightleftharpoons[\delta_{1 2} + \beta_1]
{\delta_{1 1} + \delta_{2 1}} P_1,
$$
which induces a simple birth-death 
stochastic process (again, implicitly assuming positive rate coefficients).
Network~{\rm (\ref{eq:fastslow1})} satisfies 
Assumptions~{\rm \ref{assumption:catalysed}} 
and{\rm ~\ref{assumption:catalysing}}.
Provided the rate coefficients  
are $\mathcal{O}(1)$ with respect to $0 < \varepsilon \ll 1$,
Assumption~{\rm \ref{assumption:fastslow}} 
is also fulfilled.
\end{example}

\section{Dynamical analysis}  \label{sec:dynamics}
In this section, we analyse the deterministic and stochastic 
models of the fast-slow network~(\ref{eq:R}), 
under three 
Assumptions~\ref{assumption:catalysed}--\ref{assumption:fastslow}, 
focusing on the long-term dynamics of 
species $\mathcal{P}$ (proteins). 
It is shown that, due to the time-scale separation and catalytic
nature of $\mathcal{G}$ (genes), one can `strip-off' the catalysts 
from the fast subnetwork, thus obtaining
the auxiliary network $\mathcal{R}_{\delta,\beta}$,
which plays a key dynamical role.

\subsection{Deterministic analysis} \label{sec:detanalysis}
Let us denote the concentration of species 
$\mathcal{P} = (P_1, P_2, \ldots, P_m)$ by
$\mathbf{x} = (x_1, x_2, \ldots, x_m) \in \mathbb{R}_{\ge}^m$, 
and of $\mathcal{G} = (G_1, G_2, \ldots, G_n)$ by 
$\mathbf{y} = (y_1, y_2, \ldots, y_n) \in \mathbb{R}_{\ge}^n$.
The RREs induced by~(\ref{eq:R})
(see also Section~\ref{sec:deterministic}) 
may be written as follows
\begin{align}
\varepsilon \frac{\mathrm{d} \mathbf{x}}{\mathrm{d} \tau} 
& = 
\left(\sum_{i = 1}^n 
y_i \,
\mathbf{f}_{\alpha_i} (\mathbf{x}; \, \boldsymbol{\alpha}_i) \right)
+ \mathbf{f}_{\beta}(\mathbf{x}; \boldsymbol{\beta}), \label{eq:RREa1} \\
\frac{\mathrm{d} \mathbf{y}}{\mathrm{d} \tau} & = 
\mathbf{f}_{\gamma}(\mathbf{y}; \boldsymbol{\gamma}), \label{eq:RREa2}
\end{align}
where $\tau = \varepsilon t$ is the slow time-scale, 
with $t$ being the original time-variable.
Terms $\{y_i \mathbf{f}_{\alpha_i} 
(\mathbf{x}; \, \boldsymbol{\alpha}_i)\}_{i=1}^n$ on the 
RHS of~(\ref{eq:RREa1})
arise from the catalysed networks 
$\{\mathcal{R}_{\alpha_i}(\mathcal{P})\}_{i=1}^n$
of the form~(\ref{eq:R0ai}),
while $\mathbf{f}_{\beta}(\mathbf{x}; \boldsymbol{\beta})$
arises from the the uncatalysed network 
$\mathcal{R}_{\beta}(\mathcal{P})$.
The RHS of~(\ref{eq:RREa2}) is induced by the catalysing network 
$\mathcal{R}_{\gamma}(\mathcal{G})$ 
(obtained by setting $\varepsilon = 1$
in $\mathcal{R}_{\gamma}(\mathcal{G})^{\varepsilon}$,
which is given by~(\ref{eq:R1unreg})). Let us now consider
the equilibrium behaviour of system~(\ref{eq:RREa1})--(\ref{eq:RREa2}).

By Assumption~\ref{assumption:catalysing}, 
the catalysing network is zero-deficient and 
weakly-reversible. Thus, the results
presented in Section~\ref{sec:deterministic} 
(and Theorem~\ref{theorem:comb}, in particular)
imply that equation~(\ref{eq:RREa2})
has a unique equilibrium for each initial condition, 
with the equilibrium being positive, stable and complex-balanced,
and denoted by
\begin{equation}
\mathbf{y}^*(\boldsymbol{\gamma}) 
= 
(
y_1^*(\boldsymbol{\gamma}), 
y_2^*(\boldsymbol{\gamma}), 
\ldots, 
y_n^*(\boldsymbol{\gamma}))
\in \mathbb{R}_{>}^n. 
\label{eq:CPy}
\end{equation}
The equilibria of equation~(\ref{eq:RREa1}), denoted
$\mathbf{x}^* = 
\mathbf{x}^*(y_1^*(\boldsymbol{\gamma}) \boldsymbol{\alpha}_1, 
y_2^*(\boldsymbol{\gamma}) \boldsymbol{\alpha}_2, 
\ldots, y_n^*(\boldsymbol{\gamma}) \boldsymbol{\alpha}_n, 
\boldsymbol{\beta})$, satisfy
\begin{align}
\left(\sum_{i = 1}^n 
\mathbf{f}_{\delta_i} (\mathbf{x}^*; \, y_i^* \boldsymbol{\alpha}_i) \right)
+ \mathbf{f}_{\beta}(\mathbf{x}^*; \boldsymbol{\beta}) & = 0. \label{eq:CPx}
\end{align}
Note that equation~(\ref{eq:RREa1}) 
may display attractors such as stable limit cycles, 
in which case the equilibria satisfying~(\ref{eq:CPx})
may still be relevant in providing dynamical information.
In~(\ref{eq:CPx}), we use the fact that
$y_i \, \mathbf{f}_{\alpha_i} (\mathbf{x}; \, \boldsymbol{\alpha}_i)
= \mathbf{f}_{\delta_i} (\mathbf{x}; \, y_i \boldsymbol{\alpha}_i)$,
for each fixed $\mathbf{y}$.
In particular, for $\mathbf{y} = \mathbf{y}^*$, 
 the catalysed network $\mathcal{R}_{\alpha_i}$ is the 
decatalysed network $\mathcal{R}_{\delta_i}$ with
rate coefficients $\boldsymbol{\delta}_i = y_i^* \boldsymbol{\alpha}_i$.
Thus, it follows from~(\ref{eq:CPx}) that the equilibrium of the
species of interest $\mathcal{P}$,
appearing in the composite fast-slow network~(\ref{eq:R}),
is determined by the equilibrium of the 
underlying auxiliary network 
$\mathcal{R}_{\delta,\beta}$ with
rate coefficients $\boldsymbol{\delta} = 
(y_1^*(\boldsymbol{\gamma}) \boldsymbol{\alpha}_1, 
y_2^*(\boldsymbol{\gamma}) \boldsymbol{\alpha}_2, 
\ldots, y_n^*(\boldsymbol{\gamma}) \boldsymbol{\alpha}_n)$, 
i.e. with rate coefficients $\boldsymbol{\alpha}_i$ 
each weighted by the underlying catalyst 
equilibrium $y_i^*$ given in~(\ref{eq:CPy}).
The following lemma can be deduced from
equation~(\ref{eq:CPx}).

\begin{lemma} \label{lemma:unistability}
Consider network~{\rm (\ref{eq:R})}, 
under three 
Assumptions~{\rm \ref{assumption:catalysed}}--{\rm \ref{assumption:fastslow}},
with $\mathrm{supp}(\boldsymbol{\alpha})$,
$\mathrm{supp}(\boldsymbol{\beta})$, and 
$\mathrm{supp}(\boldsymbol{\gamma})$ \emph{fixed}.
Furthermore, assume the auxiliary network 
$\mathcal{R}_{\delta,\beta}$, given by~{\rm (\ref{eq:Rdb})},
is weakly-reversible and zero-deficient 
when $\mathrm{supp}(\boldsymbol{\delta}) 
= \mathrm{supp}(\boldsymbol{\alpha})$.
Then, the {\rm RREs} underlying network~{\rm (\ref{eq:R})}
have a unique stable equilibrium 
$(\mathbf{x}^*,\mathbf{y}^*) \in \mathbb{R}_{>}^{m+n}$ for
any choice of the rate coefficients, 
i.e. network~{\rm (\ref{eq:R})} is 
unconditionally deterministically \emph{unistable}.
\end{lemma}

\noindent
Note that if $\mathcal{R}_{\delta,\beta}$
is a first-order reaction network, 
the RREs underlying network~{\rm (\ref{eq:R})}
are also deterministically unistable.

\begin{example} \label{ex:fastslownet2}
Let us consider again network~{\rm (\ref{eq:fastslow1})}
given in Example~{\rm \ref{ex:fastslownet}}.
Since the underlying auxiliary network, given by~{\rm (\ref{eq:decat1})}, 
is reversible and zero-deficient, it follows from 
Lemma~{\rm \ref{lemma:unistability}} 
that~{\rm (\ref{eq:fastslow1})}, 
with all the rate coefficients positive, is always 
deterministically unistable. Note that the composite fast-slow
network~{\rm (\ref{eq:fastslow1})} itself is \emph{not} 
zero-deficient (nor weakly-reversible). 
The same conclusion follows from the fact that~{\rm (\ref{eq:decat1})}
is an ergodic first-order network.
The underlying {\rm RREs} are given by
\begin{align}
\frac{\mathrm{d} x_1}{\mathrm{d} t} & = 
\left(y_1 \alpha_{1 1} - y_1 \alpha_{1 2} x_1 + (M - y_1) \alpha_{2 1} \right) 
- \beta_1 x_1, \label{eq:RREb1} \\
\frac{\mathrm{d} y_1}{\mathrm{d} t} & = 
- \varepsilon \gamma_{1 2} y_1 
+ \varepsilon \gamma_{2 1}(M - y_1), \label{eq:RREb2}
\end{align}
where the catalysts satisfy the 
conservation law $y_1(t) + y_2(t) = M$, for $t \ge 0$,
with $M \in \mathbb{R}_{>}$, while the equilibrium reads
\begin{equation}
(x_1^*,y_1^*) = 
\left(
\frac{y_1^* \alpha_{1 1} + (M - y_1^*) \alpha_{2 1}}
{y_1^* \alpha_{1 2} + \beta_1}, 
\frac{\gamma_{2 1}}{\gamma_{1 2} + \gamma_{2 1}} M
\right).
\label{eq:CP22} 
\end{equation}
\end{example}

\noindent
In Figure~{\rm \ref{fig:mixing1}(a)} and {\rm \ref{fig:mixing1}(c)}, 
we present in red
the $x_1$-solutions of~{\rm (\ref{eq:RREb1})}--{\rm (\ref{eq:RREb2})} 
with the catalyst
conservation constants $M = 1$ and $M = 4$, respectively, 
and
$\alpha_{1 1} = 100$, $\alpha_{1 2} = 2$,
$\alpha_{2 1} = 500$, $\beta_1= 2$, 
$\gamma_{1 2} = \gamma_{2 1} = 1$,
$\varepsilon = 10^{-2}$.
One can notice that $x_1$ approaches the 
equilibrium $x_1^* = 100$ in Figure~{\rm \ref{fig:mixing1}(a)}, 
while $x_1^* = 200$ in Figure~{\rm \ref{fig:mixing1}(c)}.
In Figures~{\rm \ref{fig:mixing1}(a)} 
and~{\rm \ref{fig:mixing1}(c)}, 
we take the catalyst initial conditions 
$(y_1(0),y_2(0)) = (0,1)$ and 
$(y_1(0),y_2(0)) = (3,1)$,
respectively. 
One can notice that, on the fast time-scale (transient dynamics),
i.e. when $\varepsilon t \ll 1$, 
$x_1$ approaches the quasi-equilibria approximately 
obtained by taking $y_1^* = y_1(0)$
in the auxiliary network, which are given by $250$
and $100$ for Figures~{\rm \ref{fig:mixing1}(a)} 
and~{\rm \ref{fig:mixing1}(c)}, respectively. 
In the next section, it is shown that auxiliary networks
with such catalyst values play an important role
in the equilibrium stochastic dynamics.

\subsection{Stochastic analysis} \label{sec:stochanalysis}
With a slight abuse of notation, we also use
$\mathbf{x} = (x_1, x_2, \ldots, x_m) \in \mathbb{Z}_{\ge}^m$
(resp. $\mathbf{y} = (y_1, y_2, \ldots, y_n) \in \mathbb{Z}_{\ge}^n$)
to denote the copy-number values of species 
$\mathcal{P} = (P_1, P_2, \ldots, P_m)$  
(resp. $\mathcal{G} = (G_1, G_2, \ldots, G_n)$).
The CME induced by~(\ref{eq:R})
 (see also Section~\ref{sec:stochastic}) is given by
\begin{equation}
\frac{\partial}{\partial \tau} p(\mathbf{x},\mathbf{y},\tau) 
= 
\left(
\frac{1}{\varepsilon} \mathcal{L}_{\alpha,\beta} + \mathcal{L}_{\gamma}
\right) 
p(\mathbf{x},\mathbf{y},\tau).
 \label{eq:CME}
\end{equation}
Operators $\mathcal{L}_{\alpha,\beta}$, and $\mathcal{L}_{\gamma}$, 
are the forward operators of networks 
$\mathcal{R}_{\alpha,\beta} = \cup_{i = 1}^n 
\mathcal{R}_{\alpha_i} \cup \mathcal{R}_{\beta}$,
and $\mathcal{R}_{\gamma}$ (obtained by setting $\varepsilon = 1$
in $\mathcal{R}_{\gamma}^{\varepsilon}$), from~(\ref{eq:R})--(\ref{eq:Rab}), 
respectively, with
\begin{equation}
\mathcal{L}_{\alpha,\beta} = 
\left(\sum_{i = 1}^n y_i \, \mathcal{L}_{\alpha_i} \right) + 
\mathcal{L}_{\beta}, \label{eq:operators}
\end{equation}
where $y_i \, \mathcal{L}_{\alpha_i}$ is the forward 
operator of the catalysed network $\mathcal{R}_{\alpha_i}$, 
while $\mathcal{L}_{\beta}$ of the
uncatalysed network $\mathcal{R}_{\beta}$.

The forward operator from~(\ref{eq:CME})
 is singularly perturbed, and, in what follows, we apply
perturbation theory to exploit this fact~\cite{Hans3,Pavliotis}. 
Substituting the power series expansion
\begin{equation}
p(\mathbf{x},\mathbf{y},\tau) 
= 
p_0(\mathbf{x},\mathbf{y},\tau) 
+ 
\varepsilon \, p_1(\mathbf{x},\mathbf{y},\tau) 
+ 
\ldots 
+ 
\varepsilon^k \, p_k(\mathbf{x},\mathbf{y},\tau) + \ldots, \nonumber
\end{equation}
into~(\ref{eq:CME}), and equating terms of equal powers in $\varepsilon$, 
the following system of equations is obtained:
\begin{align}
\mathcal{O} \left(\frac{1}{\varepsilon} \right): \; 
\mathcal{L}_{\alpha,\beta} \,
p_0(\mathbf{x},\mathbf{y},\tau) & = 0, \label{eq:QSAa1}\\
\mathcal{O}(1): \; 
\mathcal{L}_{\alpha,\beta} \,
p_1(\mathbf{x},\mathbf{y},\tau)  & = 
- \left(\mathcal{L}_{\gamma}  - \frac{\partial}{\partial \tau} \right) 
p_0(\mathbf{x},\mathbf{y},\tau). \label{eq:QSAa2}
\end{align}
Function $p_0(\mathbf{x},\mathbf{y},\tau)$ is required to be a PMF, 
and it is called the zero-order approx-imation of 
$p(\mathbf{x},\mathbf{y},\tau)$. We use the definition
of conditional PMF to write 
$p_0(\mathbf{x},\mathbf{y},\tau) 
= 
p_0(\mathbf{x}|\mathbf{y}) \, p_0(\mathbf{y},\tau)$.
Then the zero-order approximation 
of the stationary $\mathbf{x}$-marginal PMF,
which is the main object of interest in this paper, is given by
\begin{equation}
p_0(\mathbf{x}) 
=  
\sum_{\mathbf{y} \in \pi_n^N} 
p_0(\mathbf{y}) \, p_0(\mathbf{x}| \mathbf{y}),
\label{eq:mPMF}
\end{equation}
where 
\begin{equation}
\pi_n^N = \{\mathbf{y} 
= (y_1, y_2, \ldots, y_n) \in \mathbb{Z}_{\ge}^n 
\, \Big| \, 
\sum_{i=1}^n y_i = N\} \subset \mathbb{Z}_{\ge}^n, \label{eq:pin}
\end{equation}
as defined in Section~\ref{sec:stochastic}. Note that $\pi_n^N$
 may be interpreted as the set of all
the constrained $n$-element permutations of $\{0, 1, \ldots, N\}$,
under the constraint that the elements sum up to $N$.
Let us also note that $\mathbf{y}$ is an element of $\pi_n^N$ as a consequence 
of Assumption~\ref{assumption:catalysing}, demanding that
$\mathcal{R}_{\gamma}^{\varepsilon}$ is closed (conservative). $\pi_n^N$
  is also called the reaction
  simplex for the slow dynamics~\cite{Hans1}.

\medskip

\noindent
\emph{Order $1/\varepsilon$ equation} (\ref{eq:QSAa1}). 
Since $\mathcal{L}_{\alpha,\beta}$ acts only
on $\mathbf{x}$, it follows that equation~(\ref{eq:QSAa1})  
is equivalent to 
$\mathcal{L}_{\alpha,\beta} \, p_0(\mathbf{x}|\mathbf{y}) = 0$.
For a fixed $\mathbf{y}$, analogously as in
the deterministic setting, 
$\mathcal{L}_{\alpha,\beta} = \mathcal{L}_{\delta,\beta}$
with $\boldsymbol{\delta} = 
(y_1 \boldsymbol{\alpha}_1, y_2 \boldsymbol{\alpha}_2, 
\ldots, y_n \boldsymbol{\alpha}_n)$, where $\mathcal{L}_{\delta,\beta}$
is the forward operator of the auxiliary network 
$\mathcal{R}_{\delta,\beta}$. 
By Assumption~\ref{assumption:catalysed},
the CME of the auxiliary network has a unique PMF 
for any choice of the rate coefficients
(called the auxiliary PMF),
so that we may write the solution to 
equation~(\ref{eq:QSAa1}) as
\begin{equation}
p_0(\mathbf{x}| \mathbf{y}) 
= p(\mathbf{x}; \, y_1 \boldsymbol{\alpha}_1, y_2 \boldsymbol{\alpha}_2,
\ldots, y_n \boldsymbol{\alpha}_n, \boldsymbol{\beta}), 
\label{eq:p0xy}
\end{equation}
where $p(\mathbf{x}; \, y_1 \boldsymbol{\alpha}_1, y_2 \boldsymbol{\alpha}_2, 
\ldots, y_n \boldsymbol{\alpha}_n, \boldsymbol{\beta})$
is the auxiliary PMF.

\medskip

\noindent
\emph{Order $1$ equation} (\ref{eq:QSAa2}). 
The solvability condition~\cite{Pavliotis}, 
obtained by summing equation~(\ref{eq:QSAa2}) 
over the fast variable $\mathbf{x}$,
gives the effective CME
\begin{equation}
\frac{\partial}{\partial \tau} p_0(\mathbf{y},\tau) 
= 
\mathcal{L}_{\gamma} p_0(\mathbf{y},\tau). 
\label{eq:effCME}
\end{equation}
Let us focus on the stationary PMF $p_0(\mathbf{y})$.
By Assumption~\ref{assumption:catalysing},
a unique stationary PMF $p_0(\mathbf{y})$ exists.
Furthermore, Theorem~\ref{theorem:productform}
implies that the PMF takes the multinomial
product-form~(\ref{eq:mulinomialform}):
\begin{equation}
p_0(\mathbf{y}) 
= 
N! \,
\frac{(\mathbf{y}^*(\boldsymbol{\gamma}))^{\mathbf{y}}}{\mathbf{y}!}, 
\; \; \; \forall \mathbf{y} \in \pi_n^N, 
\label{eq:p0y}
\end{equation} 
where $\mathbf{y}^*(\boldsymbol{\gamma})$ is 
the unique normalized equilibrium 
obtained by setting the deterministic conservation
constant to $M = 1$ in~(\ref{eq:CPy}).

Substituting~(\ref{eq:p0xy}) and~(\ref{eq:p0y}) into~(\ref{eq:mPMF}),
one finally obtains
\begin{equation}
p_0(\mathbf{x}) 
= 
\sum_{\mathbf{y} \in \pi_n^N} \left( N! \,
\frac{(\mathbf{y}^*(\boldsymbol{\gamma}))^{\mathbf{y}}}{\mathbf{y}!}  \right)
p(\mathbf{x}; \, y_1 \boldsymbol{\alpha}_1, y_2 \boldsymbol{\alpha}_2, 
\ldots, y_n \boldsymbol{\alpha}_n, \boldsymbol{\beta}).
\label{eq:PMF}
\end{equation}
Equation~(\ref{eq:PMF}) implies that the stationary 
$\mathbf{x}$-marginal PMF, describing the equilibrium
behaviour of the fast-species $\mathcal{P}$, 
is given by a sum of the stationary PMFs of the 
underlying auxiliary networks, with rate coefficients which depend 
on $\mathbf{y}$, i.e. on the species $\mathcal{G}$.
Furthermore, each of the auxiliary PMFs is weighted 
by a coefficient which depends on
the underlying equilibrium of the catalysts, 
$\mathbf{y}^*(\boldsymbol{\gamma})$.
Put more simply, as the subnetwork 
$\mathcal{R}_{\gamma}^{\varepsilon}$
slowly switches between the states $\mathbf{y} \in \pi_n^N$,
it mixes (forms a linear combination of) the auxiliary PMFs
of the fast subnetworks 
$\mathcal{R}_{\delta,\beta}$ with rate coefficients 
$\boldsymbol{\delta} = 
(y_1 \boldsymbol{\alpha}_1, y_2 \boldsymbol{\alpha}_2, 
\ldots, y_n \boldsymbol{\alpha}_n)$. 
As shown in Section~\ref{sec:detanalysis}, such a mixing 
does not occur at the deterministic level.
Hence, we call this stochastic phenomenon 
\emph{noise-induced mixing}. Note that the deterministic equilibria
satisfying~(\ref{eq:CPx}) may correspond 
to the single term from~(\ref{eq:PMF}) for which
 $\mathbf{y}$ is closest to
$\mathbf{y}(\boldsymbol{\gamma})^*$, i.e.
when the catalysts reside in a discrete state
closest to the corresponding continuous equilibrium.

\section{Applications} \label{sec:applications}
In this section, we apply equation~(\ref{eq:PMF})
 to investigate stochastic multimodality, 
arising as a consequence of noise-induced
mixing in systems biology. 
Firstly, fast-slow networks involving zero-deficient 
and weakly-reversible auxiliary networks are considered,
 so that the auxiliary PMFs from~(\ref{eq:PMF}) are
analytically obtainable. 
It is shown via Lemma~\ref{lemma:multistability} that the 
equilibrium deterministic and stochastic dynamics of 
such fast-slow networks deviate from each other: 
the networks are deterministically unistable, but may 
display stochastic multimodality. We derive as
Lemma~\ref{lemma:bounds} bounds between which the modes 
in the underlying stationary PMF
may occur, when the auxiliary networks are first-order 
and involve only one species. First-and second-order 
auxiliary network involving multiple species are then considered. 
We investigate cases when some stationary marginal PMF are unimodal, 
while others are multimodal. Also demonstrated is that, in 
the multiple-species case, modes of different
species are generally coupled. We highlight this with an example 
where modes of the output species simply scale with modes of the input species. 
Secondly, we design a fast-slow network with third-order auxiliary 
network involving multimodality and stochastic oscillations. 
It is demonstrated that gene-regulatory-like networks, involving
as few as three species, may display arbitrary many noisy limit cycles.

\subsection{Zero-deficient and weakly-reversible auxiliary networks}
\label{sec:examples1}
In order to gain more insight into 
noise-induced mixing, we first consider a class 
of fast-slow networks~(\ref{eq:R}) for which
can obtain the auxiliary PMFs analytically,
appearing as the $\mathbf{x}$-dependent factors in~(\ref{eq:PMF}).
In particular, we consider fast-slow networks $\mathcal{R}$
with the auxiliary networks $\mathcal{R}_{\delta,\beta}$
which are zero-deficient and weakly-reversible
for any choice of the rate coefficients $(\boldsymbol{\delta},
\boldsymbol{\beta})$, with $\mathrm{supp}(\boldsymbol{\delta})$ and
$\mathrm{supp}(\boldsymbol{\beta})$ fixed, and for which the state-space 
$\mathbb{Z}_{\ge}^m$ is irreducible.
It follows from 
Theorems~\ref{theorem:comb} and~\ref{theorem:productform}
that, in this case, the auxiliary PMFs take the Poisson 
product-form~(\ref{eq:Poissonform}), so that
equation~(\ref{eq:PMF}) becomes
\begin{equation}
p_0(\mathbf{x}) = 
\sum_{\mathbf{y} \in \pi_n^N} \left( N! \, 
\frac{(\mathbf{y}^*(\boldsymbol{\gamma}))^{\mathbf{y}}}{\mathbf{y}!} \right)
\prod_{i=1}^m \mathcal{P} \left(x_i; \, 
x_i^*(y_1 \boldsymbol{\alpha}_1, y_2 \boldsymbol{\alpha}_2, 
\ldots, y_n \boldsymbol{\alpha}_n, \boldsymbol{\beta}) \right),
\label{eq:PMFzerodef}
\end{equation}
where $\mathbf{x}^*(y_1 \boldsymbol{\alpha}_1, y_2 \boldsymbol{\alpha}_2, 
\ldots, y_n \boldsymbol{\alpha}_n, \boldsymbol{\beta}) \in \mathbb{R}_{>}^m$ 
is the underlying complex-balanced equilibrium of
the auxiliary network. 
Thus, in this special case, the PMF modes are determined by the deterministic
equilibria of the auxiliary network $\mathcal{R}_{\delta,\beta}$
with rate coefficients 
$\boldsymbol{\delta} =
(y_1 \boldsymbol{\alpha}_1, y_2 \boldsymbol{\alpha}_2, 
\ldots, y_n \boldsymbol{\alpha}_n)$, $\mathbf{y} \in \pi_{n}^N$,
while the values of the marginal PMF at the modes
by the deterministic equilibrium 
of the slow network $\mathcal{R}_{\gamma}^{\varepsilon}$.

\begin{lemma} \label{lemma:multistability}
Consider network~{\rm (\ref{eq:R})}, 
under three 
Assumptions~{\rm \ref{assumption:catalysed}}--{\rm \ref{assumption:fastslow}},
with $\mathrm{supp}(\boldsymbol{\alpha})$,
$\mathrm{supp}(\boldsymbol{\beta})$, and 
$\mathrm{supp}(\boldsymbol{\gamma})$ \emph{fixed}.
Furthermore, assume the auxiliary networks 
$\mathcal{R}_{\delta,\beta}$, given by~{\rm (\ref{eq:Rdb})},
with rate coefficients $\boldsymbol{\delta} = 
(y_1 \boldsymbol{\alpha}_1, y_2 \boldsymbol{\alpha}_2,
\ldots, y_n \boldsymbol{\alpha}_n)$,
are weakly-reversible and zero-deficient for
any choice of $\mathbf{y} = (y_1, y_2, \ldots, y_n) \in \pi_n^N$.
In this case, the zero-order approximation of the 
stationary $\mathbf{x}$-marginal {\rm PMF}, 
given by~{\rm (\ref{eq:PMFzerodef})}, has maximally $|\pi_n^N|$ modes,
where set $\pi_n^N$ is given by~{\rm (\ref{eq:pin})}.
\end{lemma}
A comparison of Lemmas~\ref{lemma:unistability} 
and~\ref{lemma:multistability} identifies a class of chemical reaction networks
which are deterministically unistable, but which may be 
stochastically multimodal. Note that when the auxiliary networks 
are not zero-deficient or weakly-reversible, the auxiliary PMFs 
may be multimodal themselves. Hence, in this more general case,
the maximum number of modes in the stationary $\mathbf{x}$-marginal PMF,
given by~(\ref{eq:PMF}),
is greater than $|\pi_n^N|$. See also Section~\ref{sec:examples2}.

\subsubsection{One-species networks}
We begin by applying result~(\ref{eq:PMFzerodef})
in the simplest scenario: fast-slow networks with one-species 
first-order auxiliary networks given by
\begin{align}
\mathcal{R}_{\delta_i}: \; & & \varnothing & 
\xrightleftharpoons[y_i \alpha_{i 2}]
{y_i \alpha_{i 1}} P_1,
\qquad i \in \{1,2,\ldots, n\}, \nonumber \\
\mathcal{R}_{\beta}: \; & & \varnothing & 
\xrightleftharpoons[\beta_2]
{\beta_1} P_1.
\label{eq:example1}
\end{align}
The stationary PMF of~(\ref{eq:example1}) is a Poissionian 
with parameter 
$x_1^* = (\sum_{i=1}^n y_i \alpha_{i 1} + \beta_1)/
(\sum_{i=1}^n y_i \alpha_{i 2} + \beta_2)$,
so that~(\ref{eq:PMFzerodef}) becomes
\begin{equation}
p_0(x_1) = 
\sum_{\mathbf{y} \in \pi_n^N} \left( N! \,
\frac{(\mathbf{y}^*(\boldsymbol{\gamma}))^{\mathbf{y}}}{\mathbf{y}!}  \right)
\mathcal{P} \!\left(x_1; \, 
\frac{\sum_{i=1}^n y_i \alpha_{i 1} + \beta_1}
{\sum_{i=1}^n y_i \alpha_{i 2} + \beta_2} \right).
\label{eq:examplePMF1}
\end{equation}
We call the parameters 
$x_1^* = (\sum_{i=1}^n y_i \alpha_{i 1} + \beta_1)/
(\sum_{i=1}^n y_i \alpha_{i 2} + \beta_2)$ 
when $\mathbf{y} \in \{N \mathbf{e}_i | i \in \{1, 2,\ldots, n\}\}$
(at the boundary of the
simplex $\sum_{i = 1}^n y_i = N$)
the \emph{outer modes}, while 
when $\mathbf{y} \in \pi_{n}^{N-1}$
(in the interior of the simplex), 
the \emph{inner modes}. 
Note that the outer mode
occurring at $\mathbf{y} = N \mathbf{e}_i$
arises from network $\mathcal{R}_{\delta_i} \cup \mathcal{R}_{\beta}$
with rate coefficients $\boldsymbol{\delta}_i = N \boldsymbol{\alpha}_i$.
Denoting the smallest and largest outer modes
of network~(\ref{eq:example1}) by
\begin{align*}
x_1^{\textrm{min}}& =  
\mathop{\textrm{min}}_{i \in \{1, 2, \ldots, n\}}
\left(\frac{N \alpha_{i 1} + \beta_1}{N \alpha_{i 2} + \beta_2} \right), 
\\
x_1^{\textrm{max}} & =  
\mathop{\textrm{max}}_{i \in \{1, 2, \ldots, n\}}
\left(\frac{N \alpha_{i 1} + \beta_1}{N \alpha_{i 2} + \beta_2} \right),
\end{align*}
one can readily prove the following lemma.

\begin{lemma} \label{lemma:bounds}
Consider network~{\rm (\ref{eq:R})}, 
under three 
Assumptions~{\rm \ref{assumption:catalysed}}--{\rm \ref{assumption:fastslow}}.
Assume the underlying auxiliary network
is given by~{\rm(\ref{eq:example1})}.
Then, the inner modes of the stationary
{\rm PMF~(\ref{eq:examplePMF1})} 
are bounded below by the smallest outer mode, $x_1^{\textrm{min}}$,
and above by the largest outer mode, $x_1^{\textrm{max}}$:
\begin{align}
x_1^{\textrm{min}} &
 <
\frac{\sum_{i=1}^n y_i \alpha_{i 1} + \beta_1}
{\sum_{i=1}^n y_i \alpha_{i 2} + \beta_2}
< 
x_1^{\textrm{max}}, 
\; \; \; \forall \mathbf{y} \in \pi_{n}^{N-1}. \label{eq:bounds}
\end{align}
\end{lemma}

\noindent
Note that if all the outer modes are identical, 
then~(\ref{eq:examplePMF1}) is unimodal. 

\begin{figure}
\centerline{
\hskip 0mm
\includegraphics[width=0.5\columnwidth]{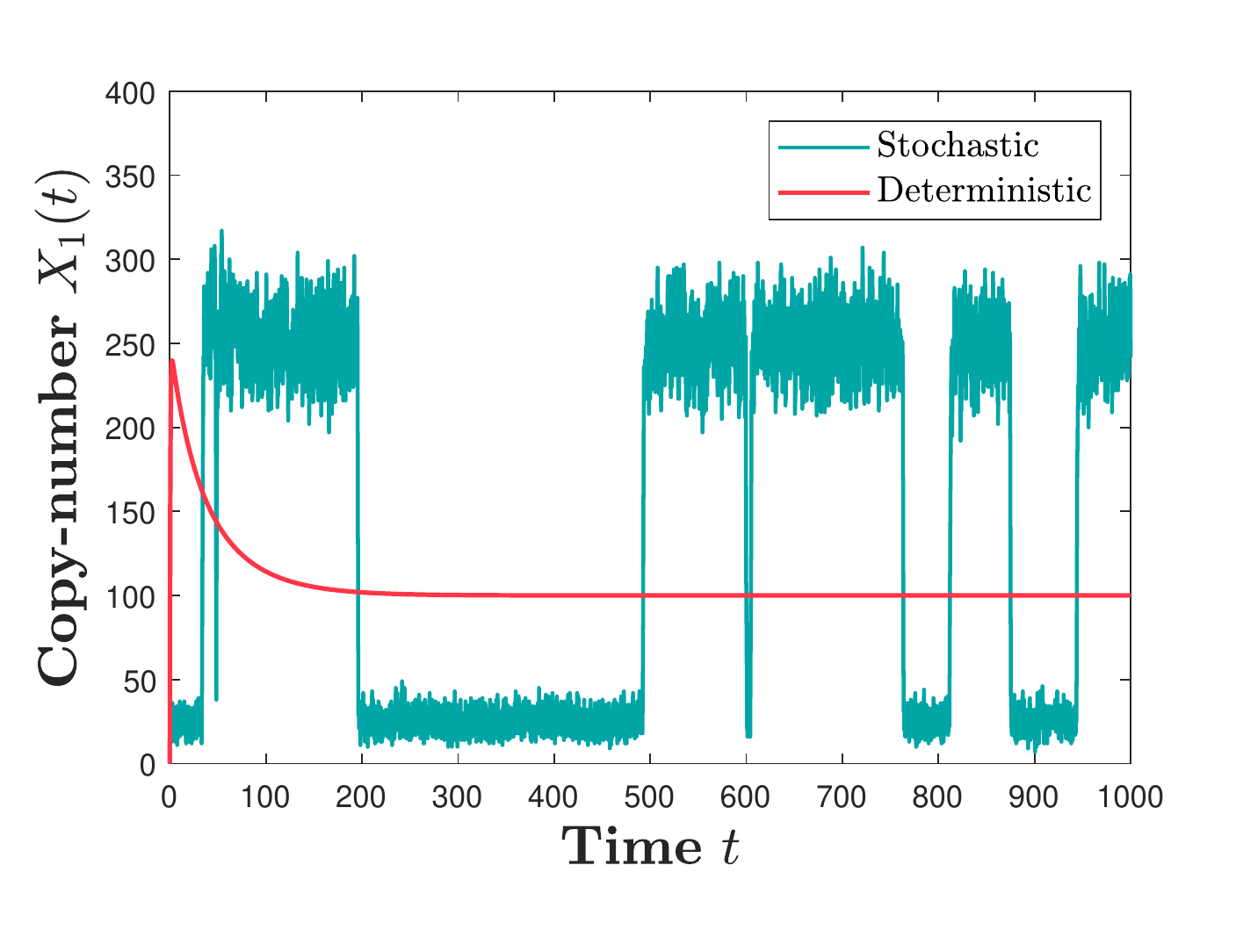}
\hskip 8mm
\includegraphics[width=0.5\columnwidth]{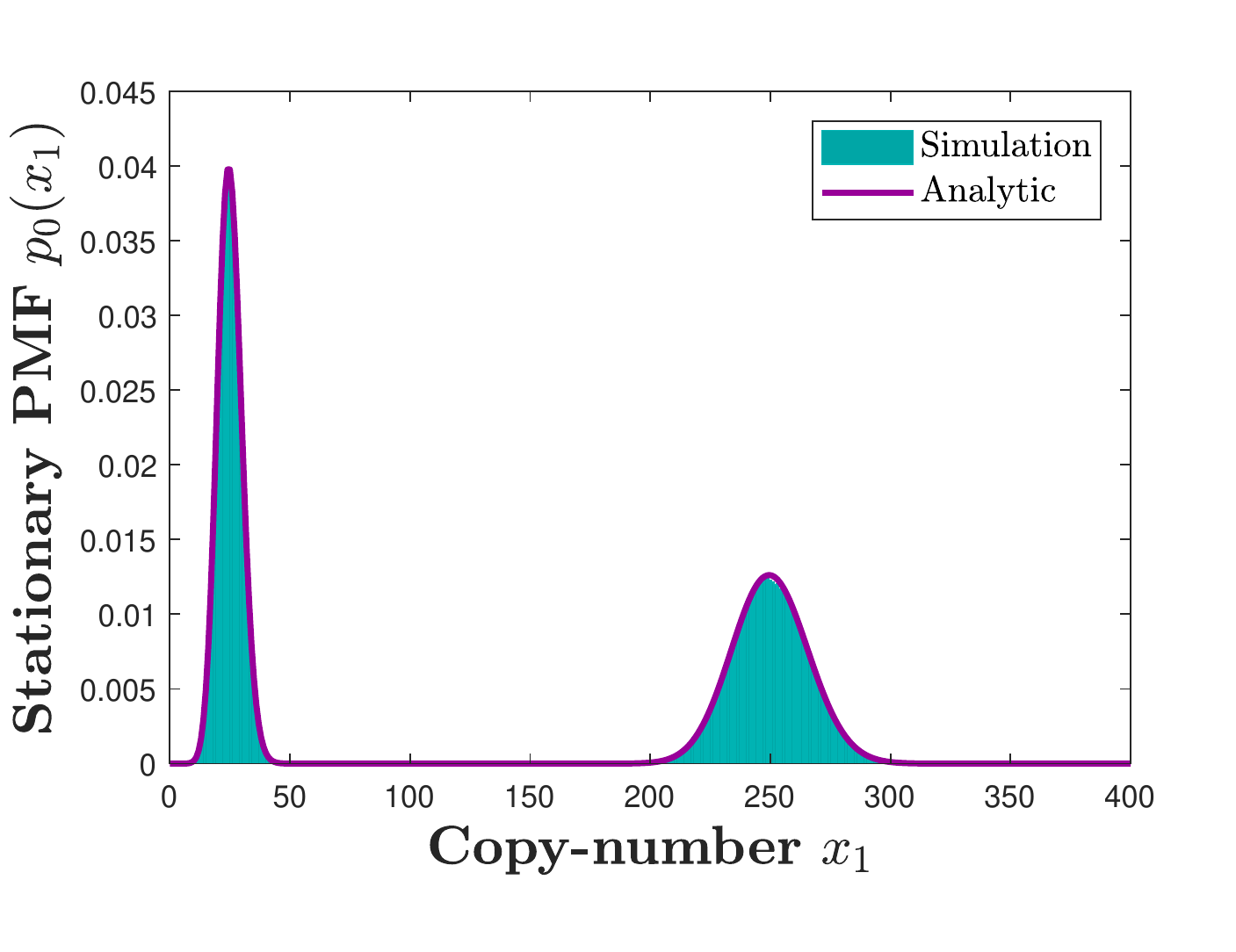}
}
\vskip -6.6cm
\leftline{\hskip 0.2cm (a) \hskip 8.5cm (b)} 
\vskip 5.5cm
\centerline{
\hskip 0mm
\includegraphics[width=0.5\columnwidth]{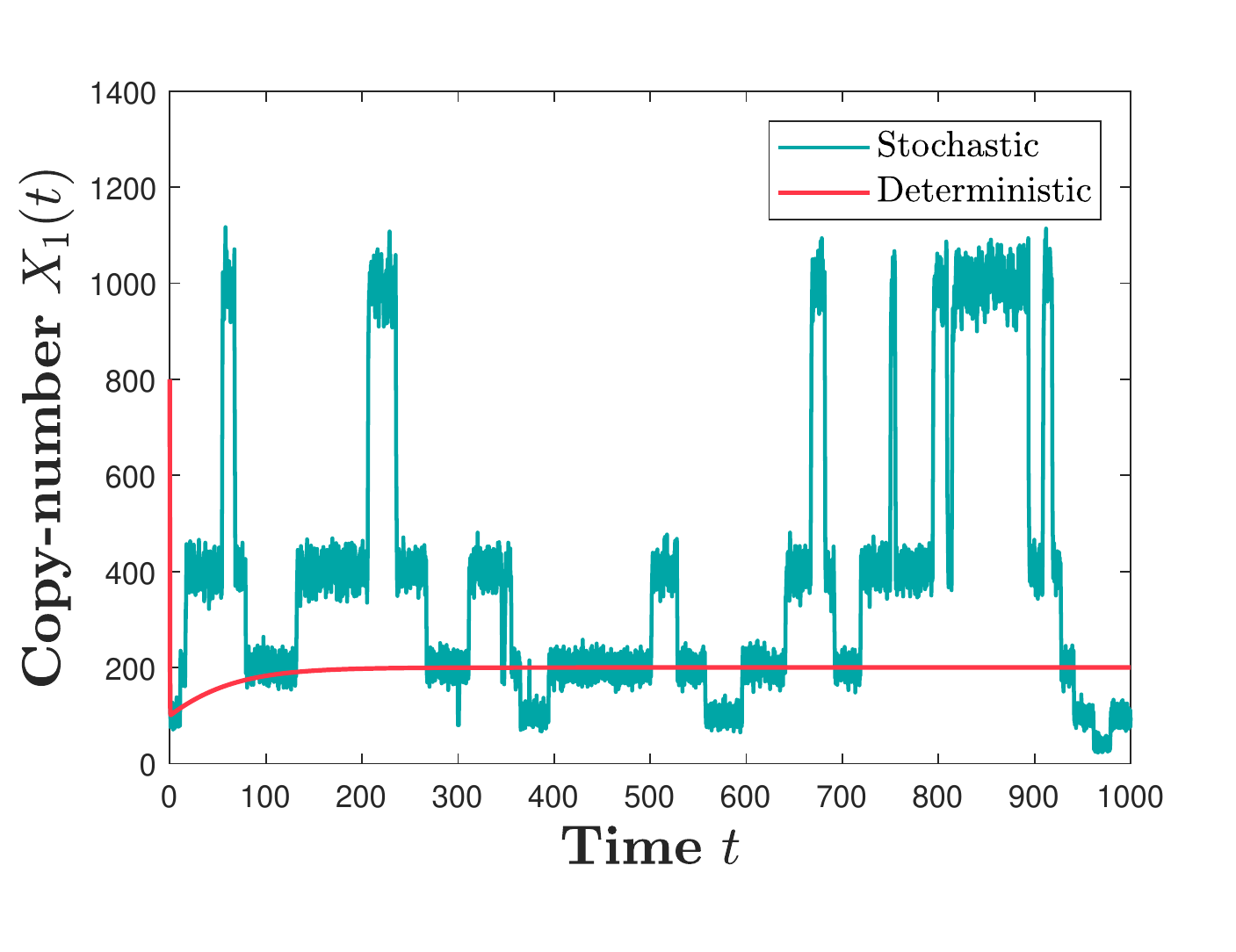}
\hskip 8mm
\includegraphics[width=0.5\columnwidth]{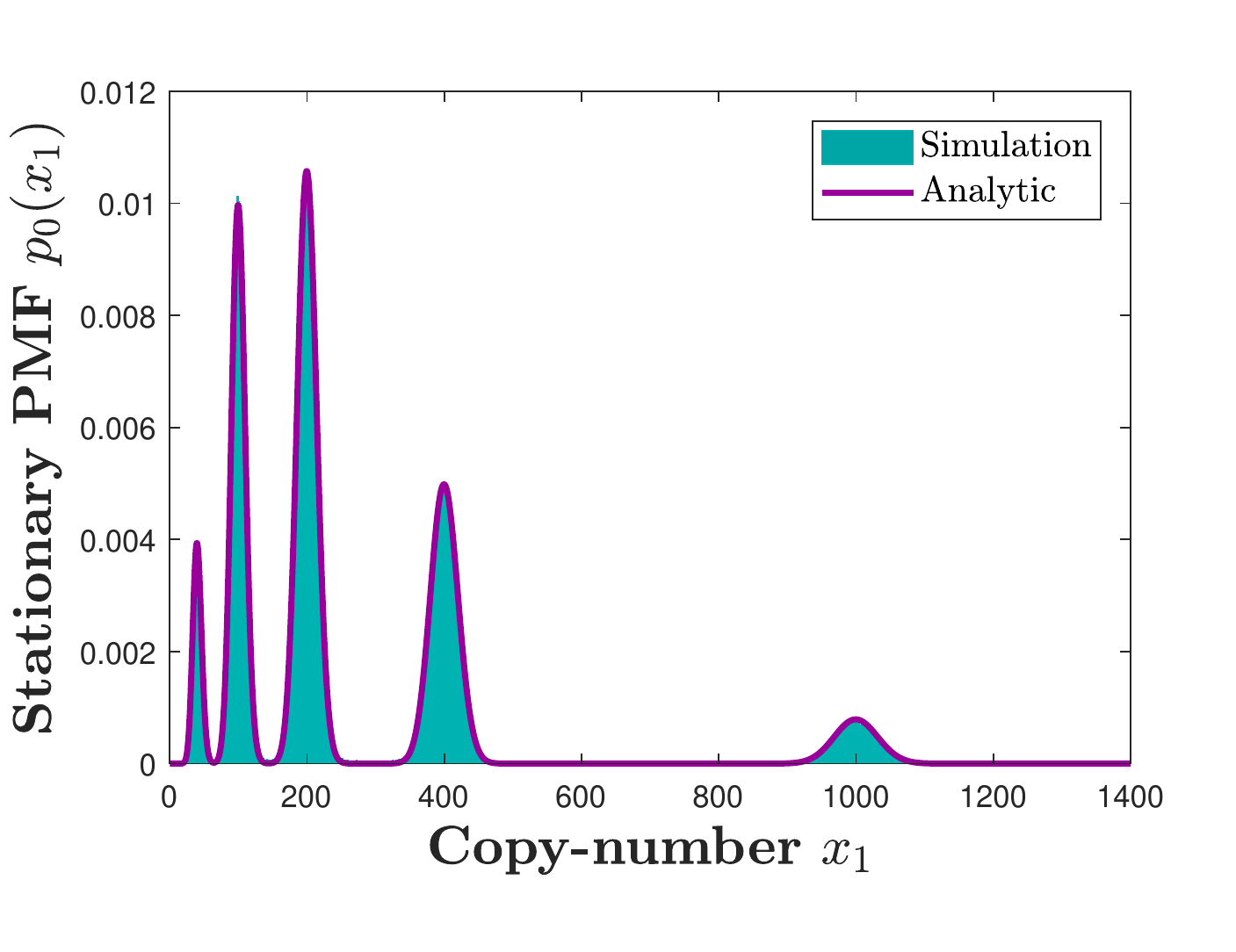}
}
\vskip -6.6cm
\leftline{\hskip 0.2cm (c) \hskip 8.5cm (d)} 
\vskip 5.5cm
\caption{ 
(a)
{\it Representative sample path for the reaction 
network~{\rm (\ref{eq:fastslow1})}, obtained 
by applying the Gillespie algorithm (blue-green), 
together with the deterministic trajectory, obtained by 
solving equations~{\rm (\ref{eq:RREb1})}--{\rm (\ref{eq:RREb2})},
in the case the two catalysts satisfy the conservation
law $y_1(t) + y_2(t) = N = M = 1$, $\forall t \ge 0$. 
The initial condition is $(x_1,y_1,y_2) = (0,0,1)$.}
(b) 
{\it Stationary $x_1$-marginal {\rm PMF} obtained by 
the Gillespie algorithm (blue-green histogram) and by
the analytic approximation~{\rm (\ref{eq:onespecies1})}
(purple).
}
(c) 
{\it Analogous results as in panel {\rm (a)} 
for the case $y_1(t) + y_2(t) = N = M = 4$, $\forall t \ge 0$, 
and with the initial condition $(x_1,y_1,y_2) = (800,3,1)$.
}
(d) 
{\it Analogous results as in panel {\rm (b)} 
for the case $y_1(t) + y_2(t) = N = M = 4$,
$\forall t \ge 0$.
The parameters are fixed to $\alpha_{1 1} = 10^2$, $\alpha_{1 2} = 2$,
$\alpha_{2 1} = 5 \times 10^2$, $\beta_1= 2$, 
$\gamma_{1 2} = \gamma_{2 1} = 1$ and
$\varepsilon = 10^{-2}$.}
}
\label{fig:mixing1}
\end{figure}

\begin{example} \label{ex:onespecies}
Let us consider again network~{\rm (\ref{eq:fastslow1})}.
The corresponding auxiliary
network~{\rm (\ref{eq:decat1})} takes the form~{\rm (\ref{eq:example1})}
with $n = 2$, $\alpha_{2 2} = \beta_1 = 0$, and with 
$\beta_2$ renamed to $\beta_1$. Fixing the conservation constant 
to $N = 1$, it follows that the possible catalyst states 
are $(y_1,y_2) \in \pi_2^1 = \{(1,0),(0,1)\}$,
and equation~{\rm (\ref{eq:examplePMF1})} becomes
\begin{equation}
p_0(x_1) 
= 
\frac{\gamma_{2 1}}{\gamma_{1 2} + \gamma_{2 1}} \,
\mathcal{P} \left(x_1; \, \frac{\alpha_{1 1}}{\alpha_{1 2} + \beta_1} \right) 
+ 
\frac{\gamma_{1 2}}{\gamma_{1 2} + \gamma_{2 1}} \,
\mathcal{P} \left(x_1; \, \frac{\alpha_{2 1}}{\beta_1} \right).
\label{eq:onespecies1}
\end{equation}
It follows from~{\rm (\ref{eq:onespecies1})} that there
are maximally two modes, which are achieved if the underlying 
two Poissonians are well-separated,
with the (outer) modes given by
\begin{equation*}
x_1^{m} \in 
\left\{\frac{\alpha_{1 1}}{\alpha_{1 2} + \beta_1}, 
\frac{\alpha_{2 1}}{\beta_1} 
\right\}. 
\end{equation*}
Let us fix the parameters to
$\alpha_{1 1} = 10^2$, $\alpha_{1 2} = 2$,
$\alpha_{2 1} = 5 \times 10^2$, $\beta_1= 2$, 
$\gamma_{1 2} = \gamma_{2 1} = 1$,
$\varepsilon = 10^{-2}$, 
as in Example~{\rm \ref{ex:fastslownet2}},
so that the two modes become 
$x_1^{m} \in \{25, 250\}$.
Note that taking $\gamma_{1 2} = \gamma_{2 1} = 1$
fixes each of the weights in~{\rm (\ref{eq:onespecies1})} to $1/2$, 
fixing the relative time the stochastic system spends
in each of the two modes. On the other hand,
taking $\varepsilon = 10^{-2}$
determines the time-scale at which the stochastic system
switches between the two modes. 
In Figure~{\rm \ref{fig:mixing1}(a)}, we display in 
blue-green a representative stochastic trajectory 
for the reaction network~{\rm (\ref{eq:fastslow1})},
obtained by applying the Gillespie stochastic simulation
algorithm~{\rm \cite{Gillespie}}.
We also show, in the same plot, the corresponding deterministic trajectory,
obtained by solving~{\rm (\ref{eq:RREb1})}--{\rm (\ref{eq:RREb2})}, 
in red. One can notice that the system is stochastically bistable,
while deterministically unistable, with the deterministic
equilibrium matching neither of the two stochastic modes.
For the gene initial condition $(y_1(0),y_2(0)) = (0,1)$, taken
in Figure~{\rm \ref{fig:mixing1}(a)}, the transient 
deterministic dynamics of $x_1$
overshoots close to the largest mode $x^m = 250$, as mentioned 
in Example~{\rm \ref{ex:fastslownet2}}.
In Figure~{\rm \ref{fig:mixing1}(b)}, we plot as the blue-green 
histogram the stationary
$x_1$-marginal {\rm PMF} obtained by utilizing the Gillespie algorithm,
while as the purple curve the analytic 
approximation~{\rm (\ref{eq:onespecies1})},
and one can see an excellent match between the two.

Fixing the conservation constant to $N = 4$,
it follows that 
$(y_1,y_2) \in \pi_2^4 = \{(4,0),(3,1)$, $(2,2),(1,3),(0,4)\}$.
Equation~{\rm (\ref{eq:examplePMF1})} then predicts
 predicts maximally $|\pi_2^4| = 5$ modes, given by
\begin{equation*}
x_1^{m} \in 
\left\{
\frac{4 \alpha_{1 1}}{4 \alpha_{1 2} + \beta_1},
\frac{3 \alpha_{1 1} + \alpha_{2 1}}{3 \alpha_{1 2} + \beta_1}, 
\frac{2 \alpha_{1 1} + 2 \alpha_{2 1}}{2 \alpha_{1 2} + \beta_1},
\frac{\alpha_{1 1} + 3 \alpha_{2 1}}{\alpha_{1 2} + \beta_1}, 
\frac{4 \alpha_{2 1}}{\beta_1}
\right\},
\end{equation*}
with $4 \alpha_{1 1}/(4 \alpha_{1 2} + \beta_1)$ 
and $4 \alpha_{2 1}/\beta_1$ 
being the outer modes,
while the rest are inner ones.
Under the same parameter choice as before,
the modes become 
$x_1^m \in 
10^2 \times
\{
0.4, 1, 2, 4, 10 
\}$.
Note that all the inner modes lie between
the two outer modes $x_1^{\textrm{min}} = 40$ 
and $x_1^{\textrm{max}} = 10^3$,
in accordance with Lemma~{\rm \ref{lemma:bounds}}.
Analogous to Figure~{\rm \ref{fig:mixing1}(a)}, 
in~Figure~{\rm \ref{fig:mixing1}(c)} 
we plot the stochastic and deterministic trajectories,
where one can notice the five stochastic modes. 
For the particular choice of the parameters, the deterministic 
equilibrium is close to the third stochastic 
mode $x_1^m = 2 \times 10^2$. 
Let us note that $(y_1(0),y_2(0)) = (3,1)$ is taken
in Figure~{\rm \ref{fig:mixing1}(c)}, and the transient dynamics of $x_1$
undershoots close to the inner mode $x_1^m = 10^2$.
In Figure~{\rm \ref{fig:mixing1}(d)}, we again demonstrate
an excellent matching between the stationary
$x_1$-marginal {\rm PMF} obtained from the 
simulations, and the one obtained from the 
analytic prediction~{\rm (\ref{eq:examplePMF1})}.
\end{example}

\subsubsection{Multiple-species networks}
When considering fast-slow networks
with multiple-species auxiliary networks,
we focus, for simplicity, on one-species marginal PMFs
(as opposed to e.g. the joint PMF).
For a given fast-slow network,
some marginal PMFs may display unimodality,
while others multimodality.
There are broadly two reasons why networks of the
form~(\ref{eq:R}) 
may display (marginal) unimodality.
Firstly, a marginal PMF may appear
unimodal if it takes a significant value at 
only one mode, i.e.
if the weights
in~(\ref{eq:PMFzerodef}) take a significant
value for only one auxiliary Poissonian.
Secondly, the $x_i$-marginal PMF $p_0(x_i)$ is unimodal
if the underlying Poissonians
$\mathcal{P} (x_i; \, x_i^*)$
from~(\ref{eq:PMFzerodef})
are not well-separated,
which occurs under insufficient separation 
between the deterministic equilibria 
$x_i^*$ $= x_i^*(y_1 \boldsymbol{\alpha}_1, y_2 \boldsymbol{\alpha}_2,
\ldots, y_n \boldsymbol{\alpha}_n, \boldsymbol{\beta})$,
$\mathbf{y} \in \pi_n^N$,
 of the auxiliary networks.
We now provide an example of the extreme case,
 when the deterministic equilibrium $x_i^*$ 
is independent of $\mathbf{y}$ (i.e. all
the deterministic equilibria of the auxiliary network coincide), so that
$p_0(x_i)$ is a sum of identical Poissonians, 
and is hence unconditionally unimodal.

\begin{figure}
\centerline{
\hskip 0mm
\includegraphics[width=0.5\columnwidth]{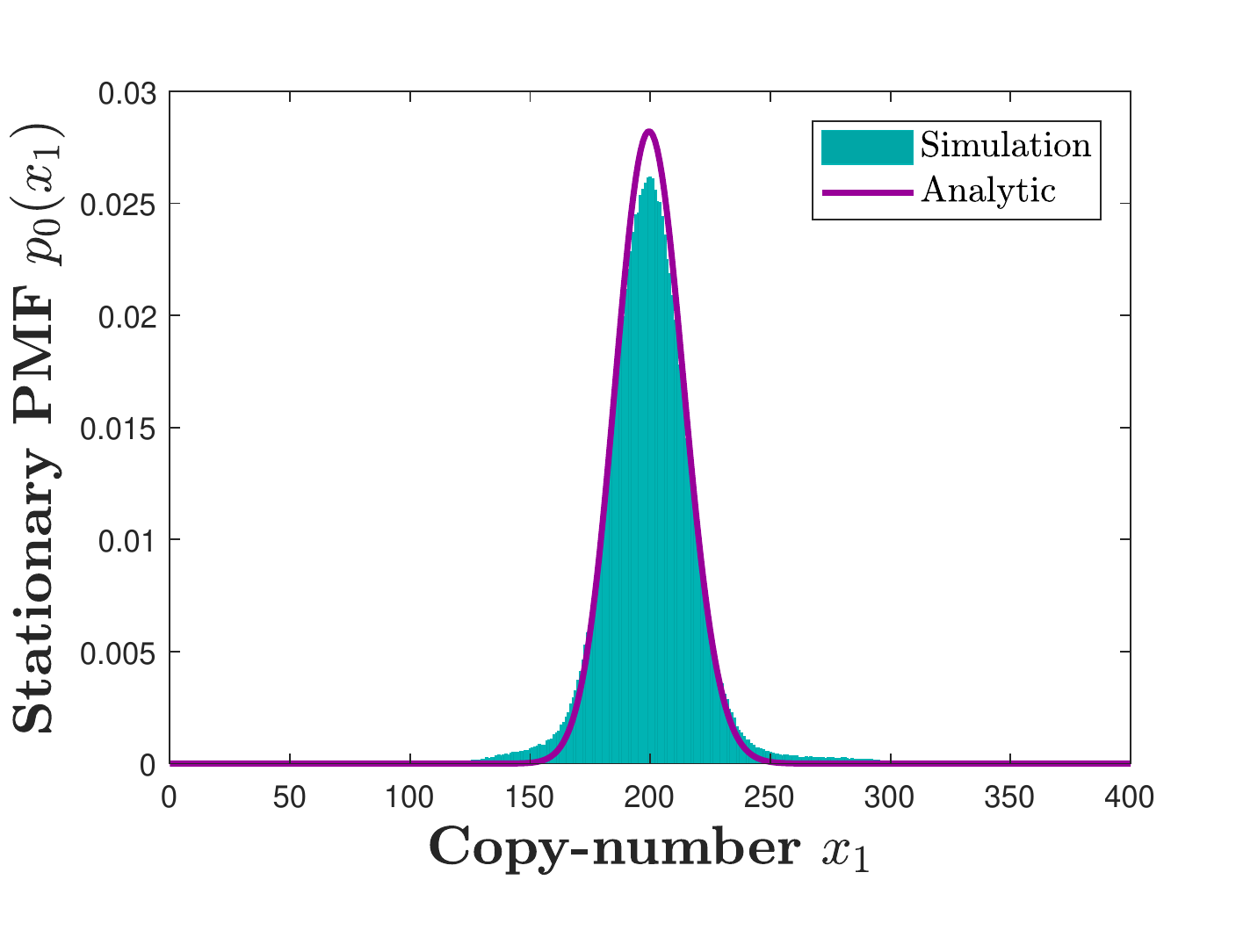}
\hskip 8mm
\includegraphics[width=0.5\columnwidth]{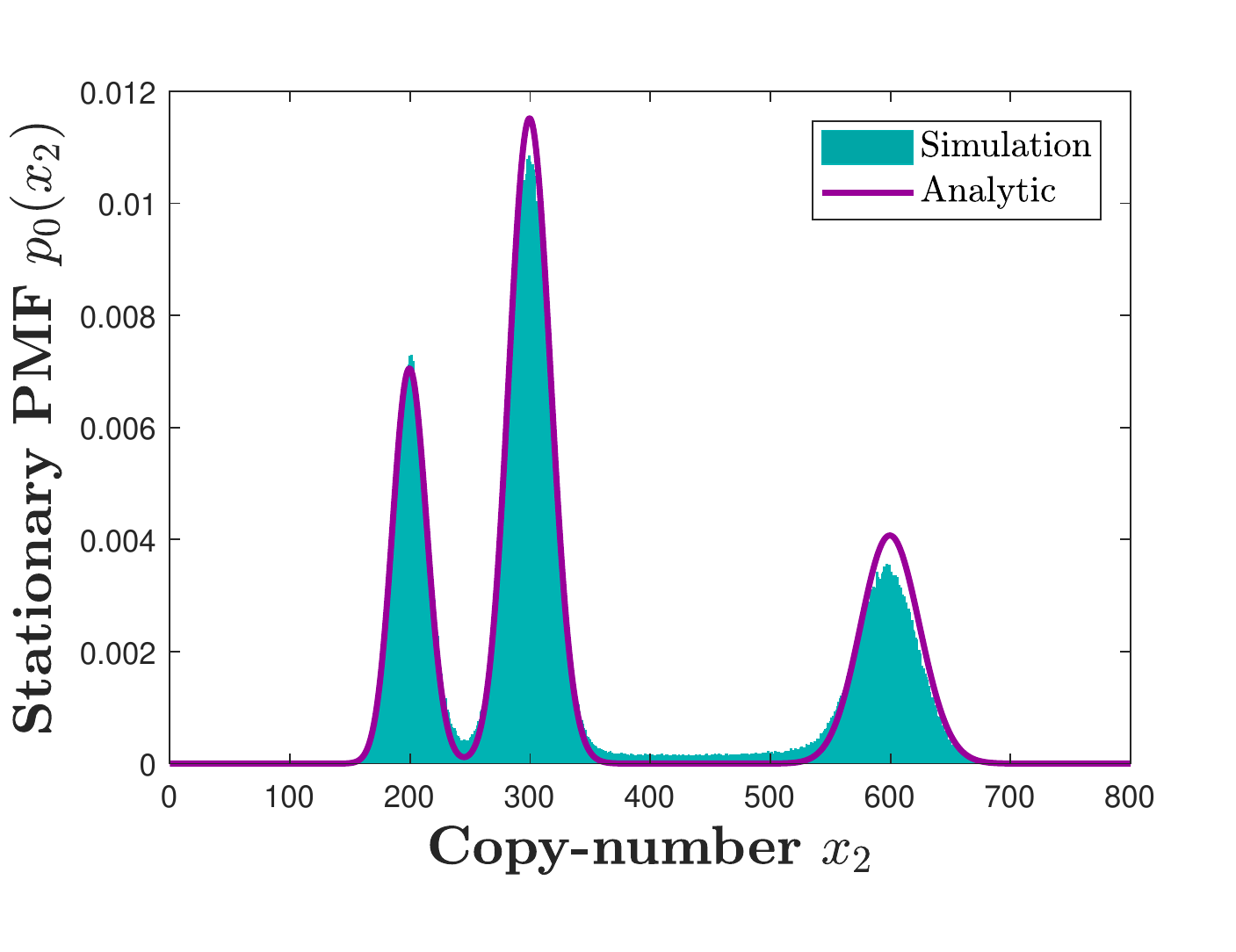}
}
\vskip -6.6cm
\leftline{\hskip 0.2cm (a) \hskip 8.5cm (b)} 
\vskip 5.5cm
\caption{ 
(a) 
{\it The stationary $x_1$--marginal {\rm PMFs} of chemical
system {\rm (\ref{eq:fastslow2})} obtained
by the Gillespie algorithm (blue-green histogram)
and by analytic approximation~{\rm (\ref{eq:PMFx1})}
(purple solid line), illustrating that species $P_1$ 
from~{\rm (\ref{eq:fastslow2})}
is unimodally distributed, with the mode $x_1^m = 200$.
}
(b) 
{\it The stationary $x_2$--marginal {\rm PMFs}  
illustrating that $P_2$ has a trimodal distribution, 
with the modes $x_2^m \in 10^2 \times \{2, 3, 6\}$.
The parameters are fixed to $\alpha_{1 1} = 1/2$,
$\alpha_{2 1} = 1/6$, $\beta_1= 1$, $\beta_2 = 2 \times 10^2$, 
$\gamma_{1 2} = \gamma_{2 1} = 1$,
$\varepsilon = 10^{-2}$ and $N= 2$.}
}
\label{fig:mixing2}
\end{figure}

\begin{example} \label{ex:twospecies}
Let us consider the following fast-slow network
\begin{align}
\mathcal{R}_{\alpha_1}: \; & & G_1 + P_2 & 
\xrightarrow[]{\alpha_{1 1}} G_1 + P_1, \nonumber \\
\mathcal{R}_{\alpha_2}: \; & & G_2 + P_2 & 
\xrightarrow[]{\alpha_{2 1}} G_2 + P_1, \nonumber \\
\mathcal{R}_{\beta}: \; & & P_1 & 
\xrightarrow[]{\beta_1} \varnothing, \nonumber \\
& & \varnothing & 
\xrightarrow[]{\beta_2} P_2, \nonumber \\
\mathcal{R}_{\gamma}^{\varepsilon}: \;
 & & G_1 & \xrightleftharpoons[\varepsilon \gamma_{2 1}]
{\varepsilon \gamma_{1 2}} G_2, 
\; \; \; \; \; 0 < \varepsilon \ll 1, \label{eq:fastslow2}  
\end{align}
involving species $\mathcal{P} = (P_1,P_2)$ and 
catalysts $\mathcal{G} = (G_1,G_2)$.
Species $P_2$ may be interpreted as the substrate needed
for the gene in both states $G_1$ and $G_2$ to build the protein $P_1$.
The auxiliary network $\mathcal{R}_{\delta,\beta}$,  
with $\boldsymbol{\delta} = 
(y_1 \alpha_{1 1}, y_2 \alpha_{2 1})$,
is given by
\begin{align}
\mathcal{R}_{\delta_1}: \; & &  P_2 & 
\xrightarrow[]{y_1 \alpha_{1 1}}  P_1, \nonumber \\
\mathcal{R}_{\delta_2}: \; & & P_2 & 
\xrightarrow[]{y_2 \alpha_{2 1}}  P_1, \nonumber \\
\mathcal{R}_{\beta}: \; & & P_1 & 
\xrightarrow[]{\beta_1} \varnothing, \nonumber \\
& & \varnothing & 
\xrightarrow[]{\beta_2} P_2. \label{eq:decat2}  
\end{align}
The deterministic equilibrium
of~{\rm (\ref{eq:decat2})} reads
\begin{equation*}
\mathbf{x}^* = 
\left(
\frac{\beta_2}{\beta_1}, 
\frac{\beta_2}{y_1 \alpha_{1 1} + y_2 \alpha_{2 1}}
\right).
\end{equation*}
In particular, $x_1^*$ is independent of the catalyst state $\mathbf{y}$.
Since~{\rm (\ref{eq:decat2})} is zero-deficient and 
weakly-reversible, equation~{\rm (\ref{eq:PMFzerodef})} is 
applicable. Summing the equation over $x_2$ and $x_1$, we
respectively obtain
\begin{align}
p_0(x_1) & =  \left(\sum_{\mathbf{y} \in \pi_2^N} N! \,
\frac{(\mathbf{y}^*(\boldsymbol{\gamma}))^{\mathbf{y}}}{\mathbf{y}!} 
 \right)
 \mathcal{P} \!\left(x_1; \, \frac{\beta_2}{\beta_1} \right), \label{eq:PMFx1}
 \\
p_0(x_2) & = \sum_{\mathbf{y} \in \pi_2^N} \left(N! \,
\frac{(\mathbf{y}^*(\boldsymbol{\gamma}))^{\mathbf{y}}}{\mathbf{y}!}  \right)
\mathcal{P} \!\left(x_2; 
\, \frac{\beta_2}{y_1 \alpha_{1 1} + y_2 \alpha_{2 1}} \right).
\label{eq:PMFx2}
\end{align}
Thus, the stationary $x_1$-marginal {\rm PMF~(\ref{eq:PMFx1})} 
is independent of $\mathbf{y}$,
and always remains unimodal. On the other hand, the stationary 
$x_2$-marginal {\rm PMF~(\ref{eq:PMFx2})} may display 
noise-induced multimodality.
Hence, the protein $P_1$ is unimodally distributed, while the 
substrate $P_2$ may be multimodally distributed.
This is also verified in Figure~{\rm \ref{fig:mixing2}} 
for a particular parameter choice, where one can also 
notice that~{\rm (\ref{eq:PMFx1})--(\ref{eq:PMFx2})}
provide an excellent approximation when $\varepsilon = 10^{-2}$.
\end{example}

\noindent
A more complicated reaction network is now presented, 
involving a second-order auxiliary network.

\begin{example} \label{ex:fourspecies}
Let us consider the following fast-slow network
\begin{align}
\mathcal{R}_{\alpha_1}: \; & & G_1 & 
\xrightarrow[]{\alpha_{1 1}} G_1 + P_1, \nonumber \\
\mathcal{R}_{\alpha_2}: \; & & G_2 & 
\xrightarrow[]{\alpha_{2 1}} G_2 + P_2, \nonumber \\
\mathcal{R}_{\alpha_3}: \; & & G_3 + P_1 & 
\xrightarrow[]{\alpha_{3 1}} G_3, \nonumber \\
& & G_3 + P_2 & 
\xrightarrow[]{\alpha_{3 2}} G_3, \nonumber \\
\mathcal{R}_{\beta}: \; 
& &  P_1 & 
\xrightarrow[]{\beta_1} \varnothing, \nonumber \\
& &  P_2 & 
\xrightarrow[]{\beta_2} \varnothing, \nonumber \\
& & P_1 + P_2 & 
\xrightleftharpoons[\beta_6]{\beta_3} P_3 
\xrightleftharpoons[\beta_5]{\beta_4} P_2 + P_4, \nonumber \\
\mathcal{R}_{\gamma}^{\varepsilon}: \;
 & & G_1 & \xrightleftharpoons[\varepsilon \gamma_{2 1}]
{\varepsilon \gamma_{1 2}} G_2
\xrightleftharpoons[\varepsilon \gamma_{3 2}]{\varepsilon \gamma_{2 3}} G_3,
\; \; \; \; \; 0 < \varepsilon \ll 1, \label{eq:fastslow3}  
\end{align}
involving species $\mathcal{P} = (P_1,P_2,P_3,P_4)$ and 
catalysts $\mathcal{G} = (G_1,G_2,G_3)$.
One may interpret $G_1,G_2,G_3$ as three possible
gene expressions: $G_1$ and $G_2$ are the producing gene states, 
creating proteins $P_1$ and $P_2$, respectively,
while $G_3$ is a degrading gene state,
destroying the two proteins.
Molecules $P_1$ and $P_2$
may also freely decay (without a direct influence of the gene), 
as well as reversibly form a complex protein $P_3$, which 
may be reversibly converted into a new protein $P_4$. 
Proteins $P_1$ and $P_2$ may be seen as input 
molecules (produced by the gene directly), while $P_3$ and $P_4$ as output
of network~{\rm (\ref{eq:fastslow3})}.
We are interested in the equilibrium dynamics of 
protein $P_4$.

\begin{figure}
\centerline{
\hskip 0mm
\includegraphics[width=0.5\columnwidth]{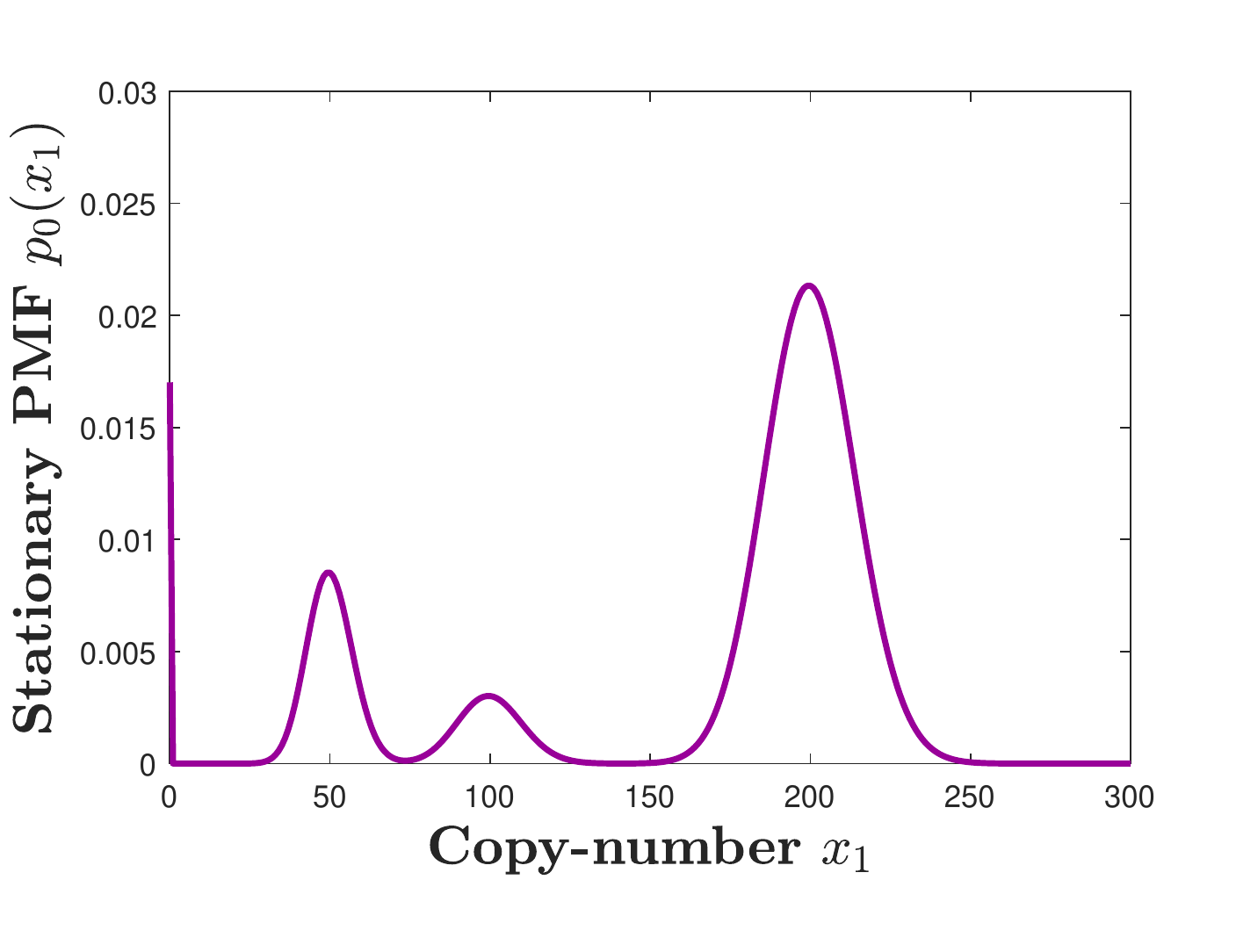}
}
\vskip -6.6cm
\leftline{\hskip 4.8cm (a)} 
\vskip 6.0cm
\centerline{
\hskip 0mm
\includegraphics[width=0.5\columnwidth]{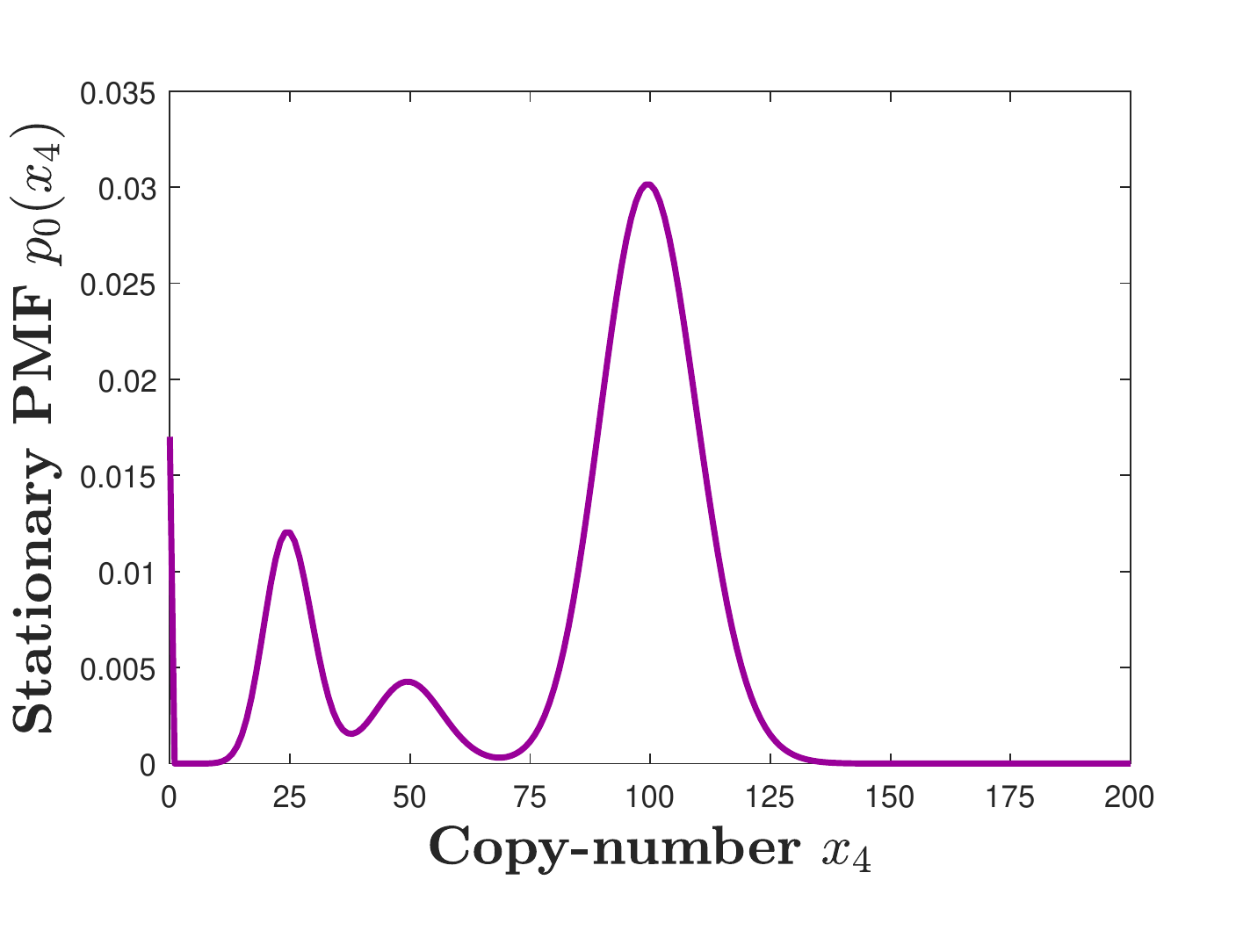}
\hskip 8mm
\includegraphics[width=0.5\columnwidth]{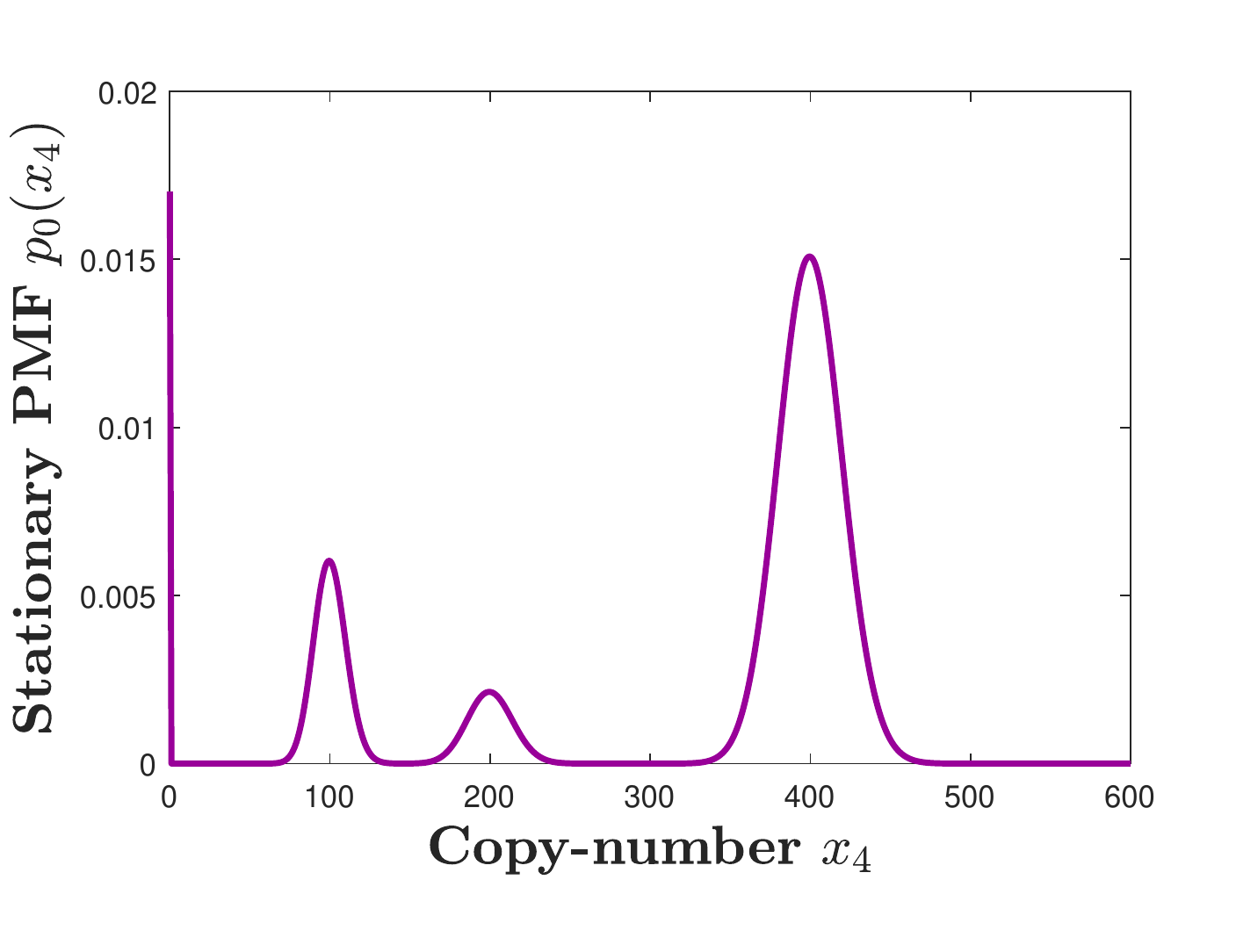}
}
\vskip -6.6cm
\leftline{\hskip 0.2cm (b)  $\frac{\beta_3 \beta_4}{\beta_5 \beta_6} = \frac{1}{2}$
 \hskip 7.0cm (c)  $\frac{\beta_3 \beta_4}{\beta_5 \beta_6} = 2$} 
\vskip 5.5cm
\caption{ 
(a) 
{\it The stationary $x_1$-marginal {\rm PMF} of system {\rm (\ref{eq:fastslow3})}
obtained by analytic approximation~{\rm (\ref{eq:PMFzerodef})}. The parameters are 
$\alpha_{1 1} = 10^2$, $\alpha_{2 1} = 50$, $\alpha_{3 1} = 1$, $\alpha_{3 2} = 1$,
$\beta_1 = \beta_2 = \beta_3 = \beta_5 = 1$, $\beta_6 = 10^2$,
$\gamma_{1 2} = \gamma_{3 2} = 1$, $\gamma_{2 1} = 20$,
$\gamma_{2 3} = 2$, $\varepsilon = 10^{-3}$, $N = 2$,
and $\beta_4 = 50$.
}
(b) 
{\it The stationary $x_4$-marginal {\rm PMF} of system {\rm (\ref{eq:fastslow3})}
given by~{\rm (\ref{eq:P34})}.} \\
(c) 
{\it The stationary $x_4$-marginal {\rm PMF} of system {\rm (\ref{eq:fastslow3})}
given by~{\rm (\ref{eq:P34})} when the value of $\beta_4$ is changed to 
$\beta_4 = 200$ (other parameters are the same as in other panels).}
}
\label{fig:mixing3}
\end{figure}

The auxiliary network $\mathcal{R}_{\delta,\beta}$,  
with $\boldsymbol{\delta} = 
(y_1 \alpha_{1 1}, y_2 \alpha_{2 1}, y_3 \alpha_{3 1}$, 
$y_3 \alpha_{3 2})$,
is given by $\mathcal{R}_{\delta,\beta}
= 
\mathcal{R}_{\delta_1} \cup \mathcal{R}_{\delta_2}
\cup \mathcal{R}_{\delta_3} \cup \mathcal{R}_{\beta},$
where $\mathcal{R}_{\beta}$ is given in 
{\rm (\ref{eq:fastslow3})} and
\begin{align*}
\mathcal{R}_{\delta_1}: \; & & \varnothing & 
\xrightarrow[]{y_1 \alpha_{1 1}} P_1,  \\
\mathcal{R}_{\delta_2}: \; & & \varnothing & 
\xrightarrow[]{y_2 \alpha_{2 1}} P_2, \\
\mathcal{R}_{\delta_3}: \; & &  P_1 & 
\xrightarrow[]{y_3 \alpha_{3 1}} \varnothing,  \\
& & P_2 & 
\xrightarrow[]{y_3 \alpha_{3 2}} \varnothing.  
\end{align*}
The deficiency of network $\mathcal{R}_{\delta,\beta}$ may be computed
using Definition~{\rm \ref{definition:deficiency}}:
$|\mathcal{C}| = 6$, $\ell = 2$ and $s = 4$, so that it is
a zero-deficient network, which is also reversible.
The deterministic equilibrium reads:
\begin{equation}
\mathbf{x}^* 
= 
\left(
\frac{y_1 \alpha_{1 1}}{y_3 \alpha_{3 1} + \beta_1},
\frac{y_2 \alpha_{2 1}}{y_3 \alpha_{3 2} + \beta_2}, 
\frac{\beta_3}{\beta_6} x_1^* x_2^*,
\frac{\beta_3 \beta_4}{\beta_5 \beta_6} x_1^*
\right).
\label{eq:det3}
\end{equation}
It follows from~{\rm (\ref{eq:PMFzerodef})} and~{\rm (\ref{eq:det3})} 
that the equilibrium behaviour of proteins $P_3$ and $P_4$,
which are produced by the gene indirectly (via $P_1$ and $P_2$),
is captured by
\begin{align}
p_0(x_3) & = 
 \sum_{\mathbf{y} \in \pi_3^N} \left(N! \,
\frac{(\mathbf{y}(\boldsymbol{\gamma})^*)^{\mathbf{y}}}{\mathbf{y}!}  \right)
 \mathcal{P} \!\left(x_3; \, \frac{\beta_3}{\beta_6} x_1^* x_2^* \right), 
\nonumber \\
p_0(x_4) & = 
 \sum_{\mathbf{y} \in \pi_3^N} \left(N! \,
\frac{(\mathbf{y}(\boldsymbol{\gamma})^*)^{\mathbf{y}}}{\mathbf{y}!}  \right)
 \mathcal{P} \!\left(x_4; \, \frac{\beta_3 \beta_4}{\beta_5 \beta_6} x_1^* \right). \label{eq:P34}
\end{align}
One can notice from~{\rm (\ref{eq:P34})} that, 
for each gene state $\mathbf{y} \in \pi_{3}^N$, 
 the mode of the complex protein $P_3$ is given by
the product of the modes of $P_1$ and $P_2$ scaled by
a factor $\beta_3/\beta_6$.
In particular, when there is only one copy of the gene, 
$N = 1$, so that $x_1^* x_2^* = 0$, it follows 
that the {\rm PMF} of $P_3$ is unimodal, and approaches the
Kronecker-delta function centered at zero
as $\varepsilon \to 0$.
On the other hand, modes of $P_4$
are modes of $P_1$ scaled by a 
factor $\beta_3 \beta_4/(\beta_5 \beta_6)$.
This is also illustrated in Figure~{\rm \ref{fig:mixing3}},
where we fix $N = 2$,
and display the stationary $x_1$-marginal {\rm PMF} in Figure~{\rm \ref{fig:mixing3}(a)}, 
while $x_4$-marginal {\rm PMF} with
$\beta_3 \beta_4/(\beta_5 \beta_6) = 1/2$
in Figure~{\rm \ref{fig:mixing3}(b)}, and with $\beta_3 \beta_4/(\beta_5 \beta_6) = 2$
in Figure~{\rm \ref{fig:mixing3}(c)}. One can notice that the modes of $p_0(x_4)$
are contracted, and dilated, by a factor of two in Figures~{\rm \ref{fig:mixing3}(b)},
and {\rm \ref{fig:mixing3}(c)}, respectively, when compared to $p_0(x_1)$. Let us note 
that, for this parameter change, only the plotted stationary $x_4$-marginal {\rm PMF} 
changes, while the other one-species marginal {\rm PMFs} remain the same, because
they are independent of $\beta_4$.
\end{example}

\subsection{Stochastic multicyclicity}
\label{sec:examples2}
In this section, we present a fast-slow network with auxiliary network
that exhibits multimodality and stochastic oscillations,
which we have constructed using~(\ref{eq:PMF}).
In this case, in contrast to Section~\ref{sec:examples1}, 
the auxiliary PMFs are not Poissonians
(more generally, Theorem~\ref{theorem:productform}
is not applicable).
The resulting fast-slow network displays an arbitrary number
of noisy limit cycles (known as stochastic multicyclicity~\cite{Me1}), 
and may illustrate the kind of stochastic dynamics arising
when a gene produces a protein whose concentration
oscillates in time.

\begin{example} \label{ex:oscillations}
Let us consider the following fast-slow network
\begin{align}
\mathcal{R}_{\alpha_1}: \; & & G_1 + 2 P_2 & 
\xrightarrow[]{\alpha_{1 1}} G_1 + P_1 + P_2, \nonumber \\
\mathcal{R}_{\beta}: \; 
& &  \varnothing & 
\xrightleftharpoons[\beta_2]{\beta_1} P_2, \nonumber \\
& &  P_2 & 
\xrightarrow[]{\beta_3} P_1, \nonumber \\
& &  P_1 + 2 P_2 & 
\xrightarrow[]{\beta_4} 3 P_2, \nonumber \\
\mathcal{R}_{\gamma}^{\varepsilon}: \;
 & & G_1 & \xrightleftharpoons[\varepsilon \gamma_{2 1}]
{\varepsilon \gamma_{1 2}} G_2,
\; \; \; \; \; 0 < \varepsilon \ll 1. \label{eq:fastslow4}  
\end{align}
Subnetwork $\mathcal{R}_{\alpha_1}$ may be seen
as a caricature of the gene, in state $G_1$,
creating products which bind two proteins
$P_2$ and then converting one of them to 
a new protein $P_1$.
Subnetwork $\mathcal{R}_{\beta}$ is the 
biochemical oscillator known as 
the Brusselator~{\rm \cite{Brusselator}}, here describing
interactions between the two proteins.

The auxiliary network $\mathcal{R}_{\delta,\beta}$,  
with $\delta_1 = y_1 \alpha_{1 1}$,
is given by $\mathcal{R}_{\delta,\beta}
= 
\mathcal{R}_{\delta_1} \cup \mathcal{R}_{\beta},$
where $\mathcal{R}_{\beta}$ is given in 
{\rm (\ref{eq:fastslow4})} and
\begin{equation}
\mathcal{R}_{\delta_1}: \; 
2 P_2  \xrightarrow[]{y_1 \alpha_{1 1}} P_1 + P_2, \label{eq:decat4}  
\end{equation}
Note that $\mathcal{R}_{\delta,\beta}$ is not zero-deficient (nor weakly-reversible),
so that~{\rm (\ref{eq:PMFzerodef})} is not applicable. We set 
$\beta_1 = \beta_2 = \beta_4 = 1$ and $\beta_3 = 10$ in the following analysis
and in Figure~{\rm \ref{fig:mixing4}}.

\smallskip

\begin{figure}
\centerline{
\hskip 0mm
\includegraphics[width=0.5\columnwidth]{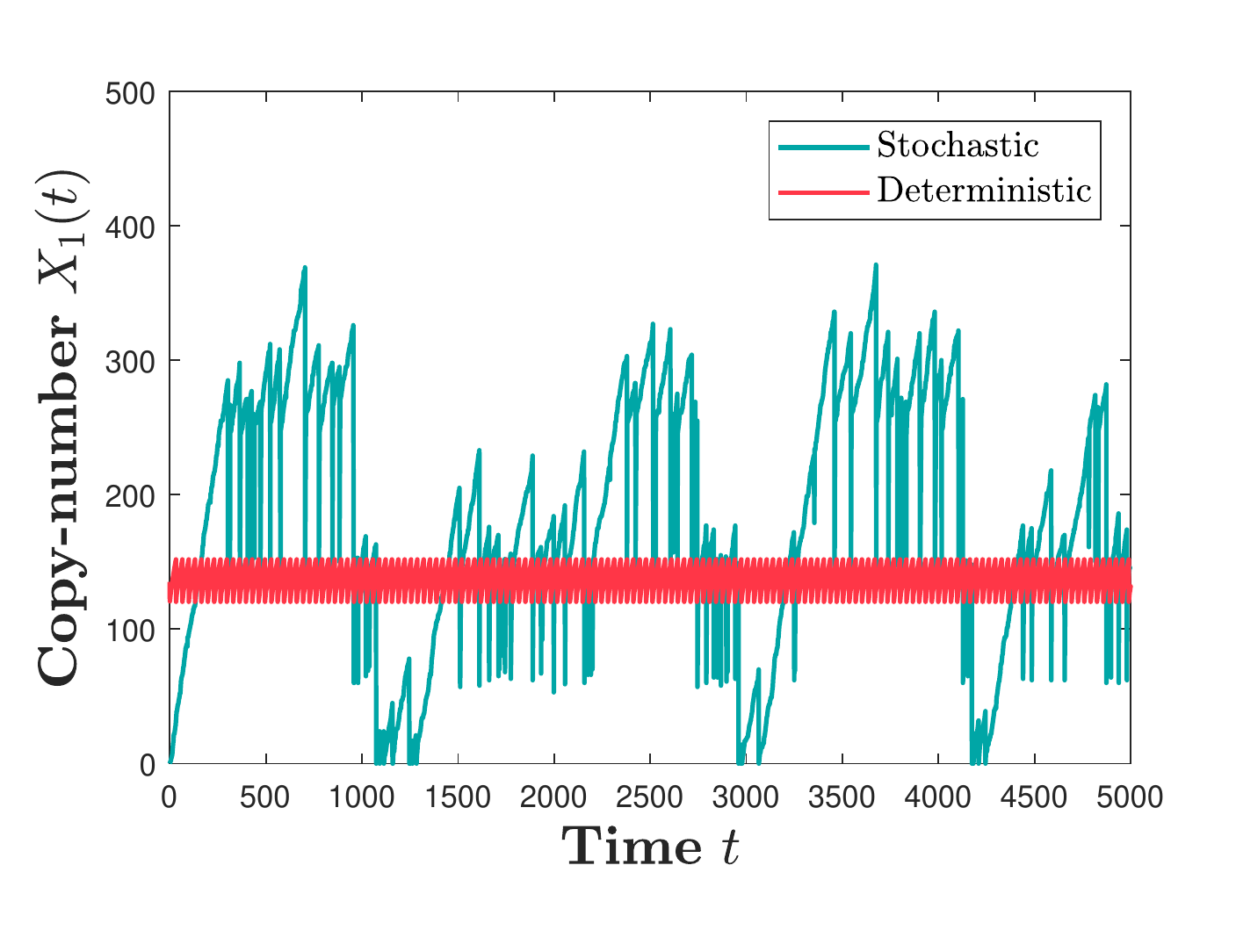}
\hskip 8mm
\includegraphics[width=0.5\columnwidth]{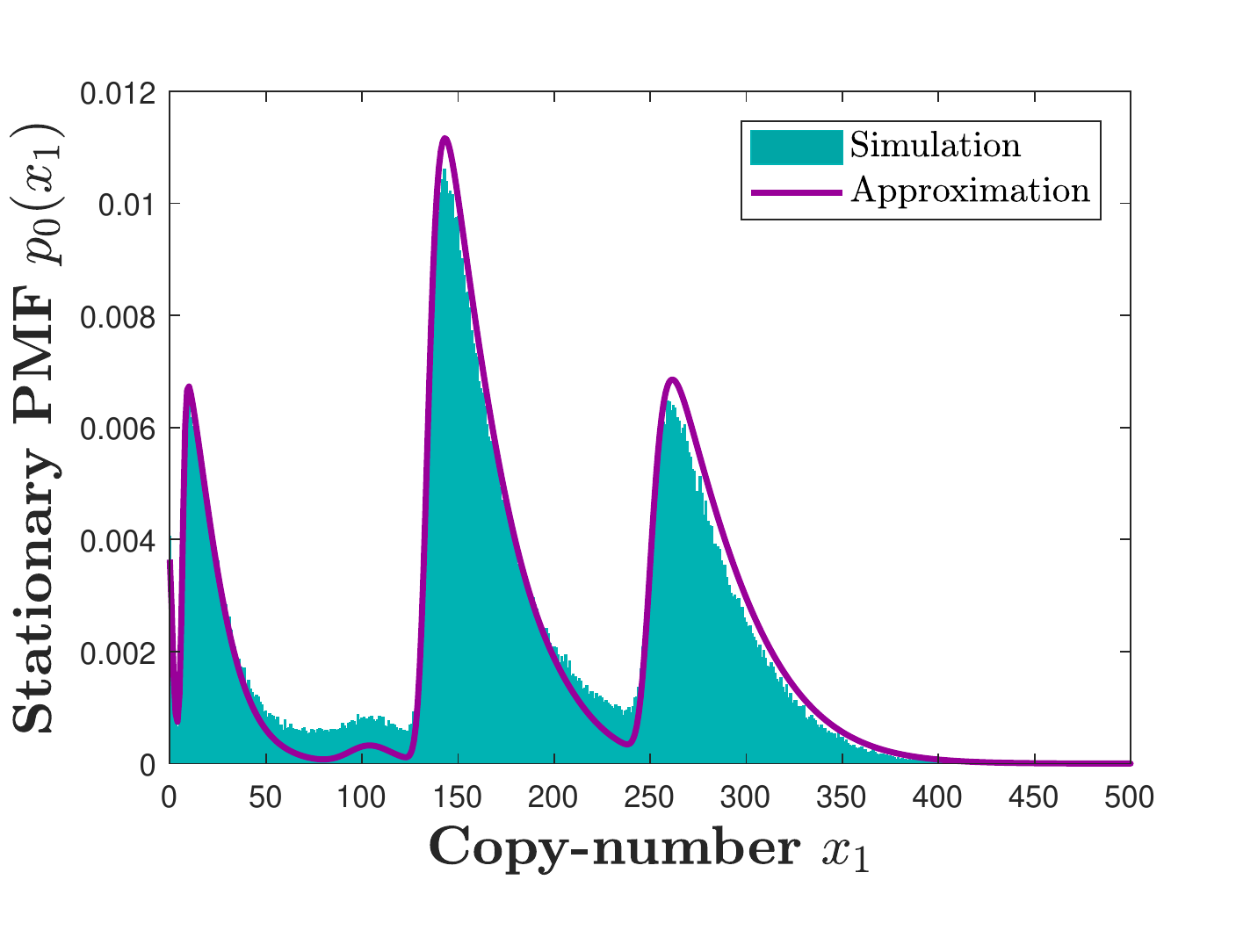}
}
\vskip -6.6cm
\leftline{\hskip 0.2cm (a) \hskip 8.5cm (b)} 
\vskip 5.5cm
\centerline{
\hskip 0mm
\includegraphics[width=0.5\columnwidth]{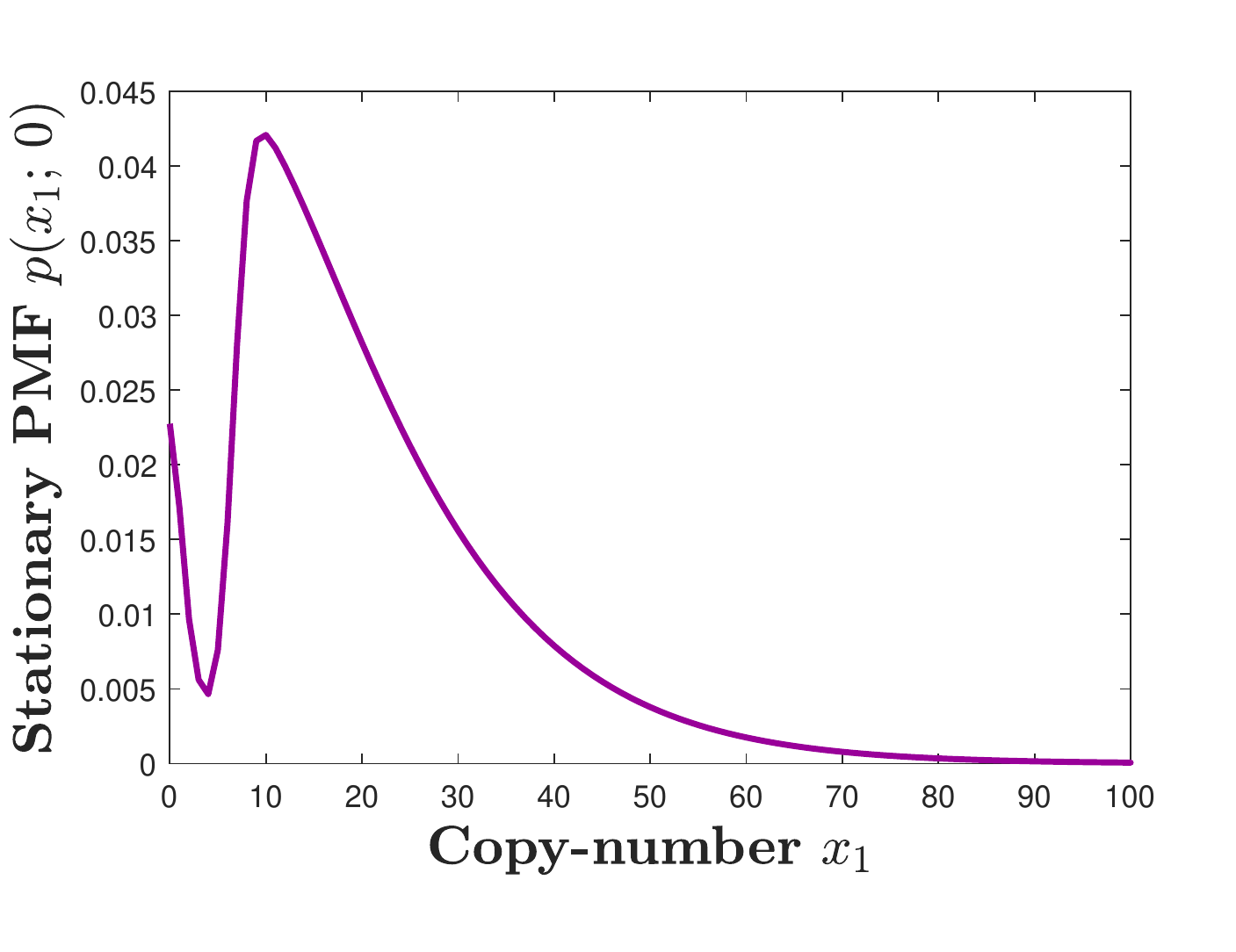}
\hskip 8mm
\includegraphics[width=0.5\columnwidth]{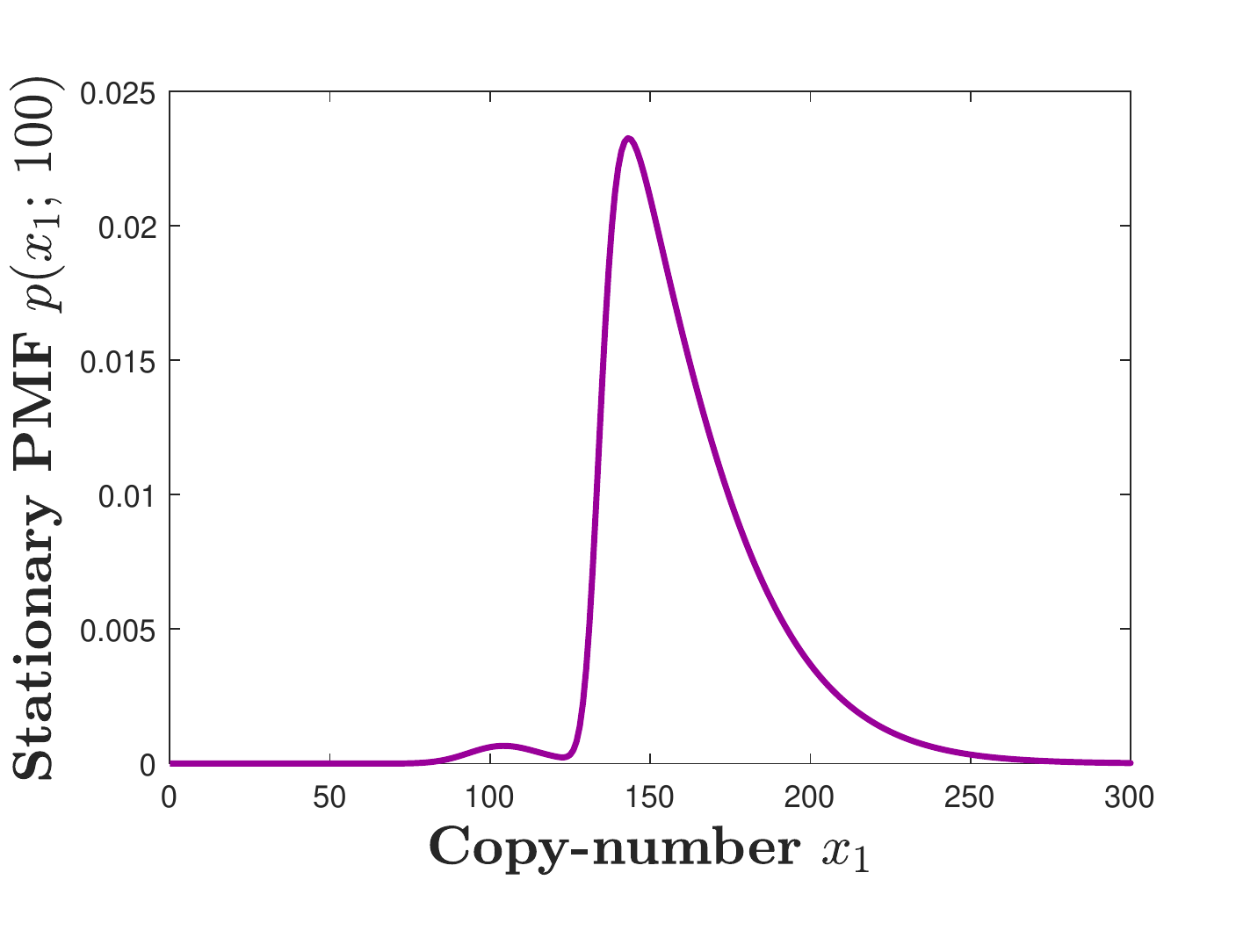}
}
\vskip -6.6cm
\leftline{\hskip 0.2cm (c) \hskip 8.5cm (d)} 
\vskip 6.0cm
\centerline{
\hskip 0mm
\includegraphics[width=0.5\columnwidth]{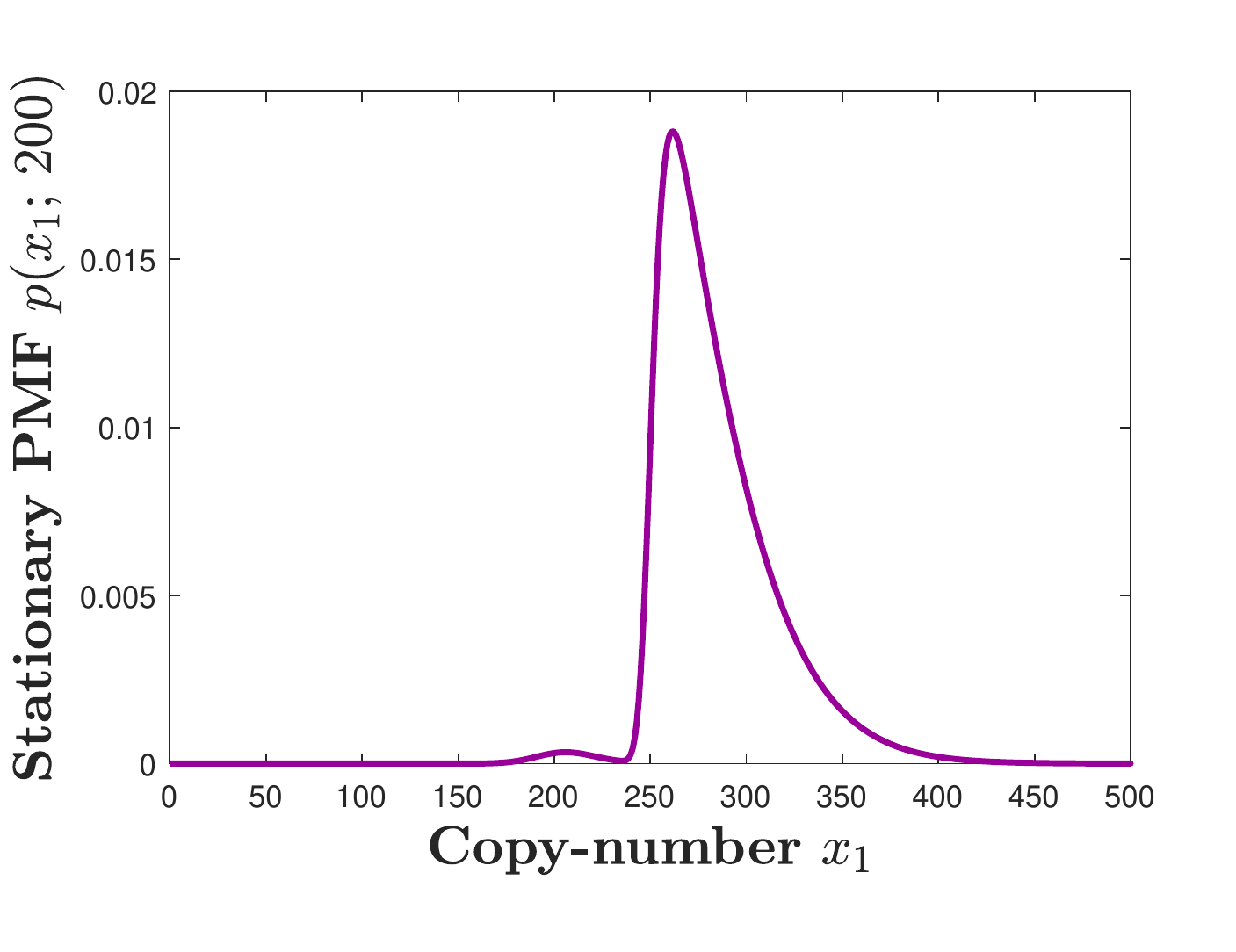}
}
\vskip -6.6cm
\leftline{\hskip 4.8cm (e)} 
\vskip 5.5cm
\caption{ 
(a)
{\it A representative sample path for network~{\rm (\ref{eq:fastslow4})} 
(blue-green), together with the deterministic trajectory, obtained by
numerically solving~{\rm (\ref{eq:RRE4})} (purple solid line).
The parameters are $\alpha_{1 1} = 10^2$, $\beta_1 = \beta_2 = \beta_4 = 1$, 
$\beta_3 = 10$, $\gamma_{1 2} = 1$, $\gamma_{2 1} = 3/2$,
$\varepsilon = 10^{-3}$, and $N= 2$.} \\
(b)
{\it The stationary $x_1$-marginal {\rm PMF} of~{\rm (\ref{eq:fastslow4})} 
(blue-green histogram) and its approximation given by~{\rm (\ref{eq:PMF14})}
(purple solid line).
} \\
(c)--(e) {\it The stationary $x_1$-marginal {\rm PMFs} of the underlying auxiliary 
network given in~{\rm (\ref{eq:PMF14})},
when}:
(c) $y_1 = 0$; (d) $y_1 = 1$; {\it and} (e) $y_2 = 2$.
}
\label{fig:mixing4}
\end{figure}

\noindent
\emph{Deterministic analysis}.The {\rm RREs} of the auxiliary network, 
with the concentration of catalyst $G_1$ set to its equilibrium value
$y_1^*$, given in~{\rm (\ref{eq:CP22})}, read
\begin{align}
\frac{\mathrm{d} x_1}{\mathrm{d} t} & = 
10 x_2 - (x_1 - y_1^* \alpha_{1 1}) \, x_2^2, \nonumber \\
\frac{\mathrm{d} x_2}{\mathrm{d} t} & = 
1 - 11 x_2 + (x_1 - y_1^* \alpha_{1 1}) \, x_2^2. \label{eq:RRE4}
\end{align}
System~{\rm (\ref{eq:RRE4})} has a unique equilibrium 
$(x_1^*,x_2^*) = (10 + y_1^* \alpha_{1 1},1)$, which 
is unstable, and surrounded by a unique stable limit cycle, for the
parameters chosen in our paper~{\rm \cite{Brusselator}}.
Deterministically, the only effect reaction $\mathcal{R}_{\delta_1}$ 
has on Brusselator $\mathcal{R}_{\beta}$ 
is to simply translate
its equilibrium and limit cycle by $y_1 \alpha_{1 1}$. 
Hence, qualitative properties of the equilibrium and limit cycle
are independent of the values of $y_1$ and $\alpha_{1 1}$.
Fixing $\alpha_{1 1} = 10^2$, the conservation constants $N = M = 2$,
and coefficients $\gamma_{1 2} = 1$, $\gamma_{2 1} = 3/2$, 
gives the equilibrium $(x_1^*,x_2^*) = (130,1)$. 
In Figure~{\rm \ref{fig:mixing4}(a)}, we show 
in red the $x_1$-solution of the {\rm RREs} 
underlying~{\rm (\ref{eq:fastslow4})} for a given initial
condition, and one can notice the time-oscillations. 

\smallskip

\noindent
\emph{Stochastic analysis}. Applying~{\rm (\ref{eq:PMF})}
on reaction network~{\rm (\ref{eq:fastslow4})},
it follows that, for sufficiently small $\varepsilon$,
the stationary $x_1$-marginal {\rm PMF}
is approximately given by
\begin{align}
p_0(x_1) 
& = 
\left(\frac{\gamma_{1 2}}{\gamma_{1 2} + \gamma_{2 1}} \right)^2 p(x_1; \, 0) 
+ 
\frac{2 \gamma_{1 2} \gamma_{2 1}}{(\gamma_{1 2} + \gamma_{2 1})^2}
\, p(x_1; \, \alpha_{1 1})
+ 
\left(\frac{\gamma_{2 1}}{\gamma_{1 2} + \gamma_{2 1}} \right)^2 p(x_1; \, 2 \alpha_{1 1}) 
\nonumber \\
& 
= 
\frac{4}{25} \, p(x_1; \, 0) 
+ 
\frac{12}{25} \, p(x_1; \, 100) 
+ 
\frac{9}{25} \, p(x_1; \, 200),
\label{eq:PMF14}
\end{align}
where $p(x_1; \, y_1 \alpha_{1 1})$ is the auxiliary {\rm PMF}.

In Figure~{\rm \ref{fig:mixing4}(a)}, we display in blue-green a representative
sample path of~{\rm (\ref{eq:fastslow4})}, which appears to switch between 
three noisy limit cycles, one of which is close to the 
deterministic limit cycle.
To gain more insight, in Figures~{\rm \ref{fig:mixing4}(c)}--{\rm \ref{fig:mixing4}(e)}, 
the auxiliary {\rm PMFs} $p(x_1; \, 0)$,
$p(x_1; \, 100)$ and $p(x_1; \, 200)$ from~{\rm (\ref{eq:PMF14})} are
presented, respectively,
obtained by numerically solving the two-species {\rm CME} 
for auxiliary network $\mathcal{R}_{\delta,\beta}
= 
\mathcal{R}_{\delta_1} \cup \mathcal{R}_{\beta}$
given in {\rm (\ref{eq:fastslow4})} and~{\rm (\ref{eq:decat4})}.
In all the three cases, the underlying deterministic model 
displays only one stable set - the limit cycle, while the 
auxiliary {\rm PMFs} are bimodal. In Figure~{\rm \ref{fig:mixing4}(b)}, 
we present as the blue-green histogram 
the $x_1$-marginal {\rm PMF} obtained from simulations, 
and as the purple curve the analytic approximation
given by the weighted sum~{\rm (\ref{eq:PMF14})}.
One can notice a good match for $\varepsilon = 10^{-3}$
taken in Figure~{\rm \ref{fig:mixing4}}.
In addition to the three modes where the {\rm PMF} takes largest
values, there are two other modes (one at $0$, 
and one near $100$), arising from $p(x_1; \, 0)$
and $p(x_1; \, 100)$. On the other hand, the second mode of
$p(x_1; \, 200)$, appearing near $200$ in Figure~{\rm \ref{fig:mixing4}(e)},
is merged with $p(x_1; \, 100)$ for the particular choice
of the parameters. Let us note that, while the
stationary $x_1$-marginal {\rm PMF} displays multimodality, 
the stationary $x_2$-marginal {\rm PMF} is unimodal and 
concentrated around $0$. This results from the fact 
that $X_2(t)$ spends most
of the time near zero for each of the three noisy limit cycles.

More generally, taking the conservation constant $N \ge 0$,
network~{\rm (\ref{eq:fastslow4})} may display $(N+1)$ distinct noisy
limit cycle. Moreover, replacing the slow subnetwork 
$$
G_1 \xrightleftharpoons[\varepsilon \gamma_{2 1}]
{\varepsilon \gamma_{1 2}} G_2, 
\qquad \mbox{from~{\rm (\ref{eq:fastslow4})} by}
\qquad
G_1 \xrightleftharpoons[\varepsilon \gamma_{0 1}]
{\varepsilon \gamma_{1 0}} \varnothing
$$ (which relaxes
Assumption~{\rm \ref{assumption:catalysing}}, see also 
Section~{\rm \ref{sec:conclusion}} for a discussion), the resulting
fast-slow network may display an \emph{infinite} number
of noisy limit cycles.
\end{example}

\section{Summary and conclusion} \label{sec:conclusion}
In this paper, we have introduced a class of
chemical reaction networks under mass-action kinetics,
involving two time-scales and catalytic species, 
and inspired by gene-regulatory networks~\cite{Kepler},
whose deterministic and stochastic
descriptions display `deviant' differences~\cite{Samoilov}.
More precisely, fast-slow networks 
of the form~(\ref{eq:R})--(\ref{eq:Rab}),
under three Assumptions~\ref{assumption:catalysed}--~\ref{assumption:fastslow},
as defined in Section~\ref{sec:probform}, have been considered.
By analyzing the underlying dynamical models
in Section~\ref{sec:dynamics}, we have
identified a novel stochastic phenomenon
causing the qualitative differences between 
the deterministic and stochastic models. 
In particular, it is shown that, 
as a result of the conversions among the catalysts (genes) in the 
slow subnetwork, the fast species
(proteins) have a probability distribution which is
a mixture of the probability distributions of 
modified fast subnetworks, called auxiliary networks,
which are obtained if the catalysts are `stripped-off'. 
We call this phenomeon
\emph{noise-induced mixing}, and 
it is captured in the central result in this paper: 
equation~(\ref{eq:PMF}), which was obtained by 
 applying first-order perturbation theory 
on the underlying singularly perturbed CME.

In Section~\ref{sec:applications}, we have applied the result
to investigate multimodality in the context
of systems biology. In Section~\ref{sec:examples1},
fast-slow reaction networks with auxiliary networks under suitable constraints
(zero-deficiency and weak-reversibility) were considered,
allowing for analytic results. 
It is shown in Lemma~\ref{lemma:multistability} that, under these constraints,
 while the deterministic model is always unistable, 
the stochastic model may display multimodality.
When the auxiliary networks involve only one species
and first-order reactions, we also derived bounds on the modes, 
given as Lemma~\ref{lemma:bounds}.
When the auxiliary networks involve multiple species, 
we discuss, and demonstrate via examples~(\ref{eq:fastslow2})
and~(\ref{eq:fastslow3}), that some species may be unimodal,
while other multimodal, and that modes of different species
are generally coupled.
In Section~\ref{sec:examples2}, a reaction network involving 
an oscillator is presented, capturing the kind of behaviour
which may arise in gene-regulatory networks involving
proteins whose concentrations oscillate in time. 
We show that, as a result of noise-induced mixing,
the reaction network may display stochastic multimodality,
where the modes correspond to copies of the underlying unique deterministic
limit cycle, thus also showing that gene-regulatory networks, involving
as few as three species, may display an arbitrary number of
noisy limit cycles.
It was also demonstrated that result~(\ref{eq:PMF})
is beneficial for numerical simulations - 
instead of simulating the higher-dimensional stiff dynamics 
of the fast-slow networks,
involving the small parameter $\varepsilon$, 
one may instead simulate the underlying lower-dimensional auxiliary networks
and use~(\ref{eq:PMF}) (see also~\cite{Hans3,Cotter,Ioannis,Cotter3} 
for discussions on simulating general fast-slow networks).

Three Assumptions~\ref{assumption:catalysed}--\ref{assumption:fastslow} have 
been made in this paper to facilitate the analysis. However, noise-induced mixing 
occurs in a broader class of reaction networks.
For example, we may relax the assumptions about 
the catalysing network, made in Assumption~\ref{assumption:catalysing}, 
in the following two ways. Firstly, we may allow the 
slow subnetwork $\mathcal{R}_{\gamma}^{\varepsilon}(\mathcal{G})$ 
to be \emph{open}, in which case the multinomial function~(\ref{eq:p0y}),
appearing as $\mathbf{x}$-independent weights in~(\ref{eq:PMF}),
is replaced with the Poissonian function of the form~(\ref{eq:Poissonform}).
Secondly, we may consider the more general \emph{regulated}
slow subnetworks, $\mathcal{R}_{\gamma}^{\varepsilon}
(\mathcal{P},\mathcal{G})$, describing gene-regulatory
networks with feedback~\cite{Kepler}. 
In this case, the derivation from Section~\ref{sec:stochanalysis}
remains valid under one modification:
the RHS of the effective CME~(\ref{eq:effCME}) depends 
on the moments of the fast species $\mathbf{x}$ (proteins)
with respect to the auxiliary PMF, which themselves depend
 on the catalysts (genes) $\mathbf{y}$.
As a consequence, the weights from~(\ref{eq:PMF})
then generally have a different form. However, 
the auxiliary PMFs ($\mathbf{x}$-dependent factors from~(\ref{eq:PMF}))
remain unchanged, so that noise-induced mixing
remains to operate. Put more simply, proteins in the
discussed gene-regulatory networks
with and without feedback have approximately the same modes, 
but the height of the probability distribution at the modes 
is generally different.
Note that networks~(\ref{eq:R})--(\ref{eq:Rab})
experience, not only long-term, but also
 \emph{transient} noise-induced mixing:
if the time-dependent PMF $p_0(\mathbf{y},\tau)$,
satisfying~(\ref{eq:effCME}), is substituted 
into~(\ref{eq:PMF}), one obtains an
approximation to the time-dependent
marginal PMF, $p_0(\mathbf{x},\tau)$,
which has the same form as~(\ref{eq:PMF}),
but with suitable time-dependent weights.

Finally, let us note that noise-induced mixing may also
be applicable to the field of synthetic biology,
which aims to design reaction systems with  
predefined behaviours~\cite{Me2}. In particular, given 
a target probability distribution, 
one may construct a suitable fast-slow network,
such that its probability distribution, given by~(\ref{eq:PMF}), 
approximates the target one.

\section{Acknowledgements}
This work was supported by NIH Grant $\#29123$ and a Visiting Research Fellowship 
from Merton College, Oxford, awarded to Hans Othmer. Radek Erban would also like
to thank the Royal Society for a University Research Fellowship.

\label{lastpage}

\end{document}